\documentclass[preprint]{elsarticle}

\usepackage{amssymb}
\usepackage{amsthm}

\usepackage{lineno}

\usepackage{xcolor}
\usepackage{multirow}
\usepackage{amsmath}
\usepackage{graphicx}

\usepackage{booktabs}
\usepackage{textcomp}
\usepackage{rotating}
\usepackage{xcolor}
\usepackage{framed}
\usepackage{nomencl}
\usepackage{verbatim}

\journal{Solar Energy}

\begin{document}

\begin{frontmatter}



\title{Practical Recommendations for the Design of Automatic Fault Detection Algorithms Based on Experiments with Field Monitoring Data}


\author[IST]{Eduardo Abdon Sarquis Filho\corref{mycorrespondingauthor}}
\ead{eduardosarquis@tecnico.ulisboa.pt}

\affiliation[IST]{organization={IDMEC, Instituto Superior Técnico},
            addressline={Av. Rovisco Pais 1}, 
            postcode={1049-001}, 
            state={Lisboa},
            country={Portugal}}

\author[enmova]{Björn Müller}
\author[fraunhofer]{Nicolas Holland}
\author[fraunhofer]{Christian Reise}
\author[fraunhofer]{Klaus Kiefer}
\author[pohlen]{Bernd Kollosch}
\author[IST]{Paulo J. Costa Branco}

\affiliation[fraunhofer]{organization={Fraunhofer Institute for Solar Energy Systems ISE},
            addressline={Heidenhofstrasse 2}, 
            postcode={79110}, 
            state={Freiburg},
            country={Germany}}
            
\affiliation[enmova]{organization={Enmova GmbH},
            addressline={Basler Str. 115}, 
            postcode={79115}, 
            state={Freiburg},
            country={Germany}}

\affiliation[pohlen]{organization={Pohlen Solar GmbH},
            addressline={Am Pannhaus 2-10}, 
            postcode={52511}, 
            state={Geilenkirchen},
            country={Germany}}

\begin{abstract}
Automatic fault detection (AFD) is a key technology to optimize the Operation and Maintenance of photovoltaic (PV) systems portfolios. 
A very common approach to detect faults in PV systems is based on the comparison between measured and simulated performance. Although this approach has been explored by many authors, due to the lack a common basis for evaluating their performance, it is still unclear what are the influencing aspects in the design of AFD algorithms. In this study, a series of AFD algorithms have been tested under real operating conditions, using monitoring data collected over 58 months on 80 rooftop-type PV systems installed in Germany. The results shown that this type of AFD algorithm have the potential to detect up to 82.8\% of the energy losses with specificity above 90\%. 
In general, the higher the simulation accuracy, the higher the specificity. The use of less accurate simulations can increase sensitivity at the cost of decreasing specificity. Analyzing the measurements individually makes the algorithm less sensitive to the simulation accuracy. The use of machine learning clustering algorithm for the statistical analysis showed exceptional ability to prevent false alerts, even in cases where the modeling accuracy is not high. If a slightly higher level of false alerts can be tolerated, the analysis of daily PR using a Shewhart chart provides the high sensitivity with an exceptionally simple solution with no need for more complex algorithms for modeling or clustering.
\end{abstract}


\begin{highlights}
\item Recommendations for designing automatic fault detection algorithms for PV systems
\item Results based on tests with 5 years monitoring data from 80 rooftop-type PV systems
\item Statistical analysis based on clustering shows very few false alerts in all cases
\item Using less accurate modeling increases sensitivity, but decreases specificity
\item Daily PR monitoring with control chart can be a reasonable and simple solution
\end{highlights}


\end{frontmatter}


\section{Introduction} \label{sec:intro}

Automatic fault detection (AFD) is a key technology to optimize Operation and Maintenance (O\&M) of photovoltaic (PV) systems portfolios. In the literature, several recent studies suggest methods and procedures for AFD \cite{madeti_comprehensive_2017,mellit_fault_2018,pillai_comparative_2019,akram_modeling_2015}. In general, the procedures analyze information from an online monitoring database in search of abnormal behavior. A report on the monitoring for fault detection in PV systems recently published \cite{rapaport2021advanced} highlights four main approaches observed in the literature, namely: (a) \textit{identification of electrical signatures}; (b) \textit{comparison of historical data to current PV system behavior}; (c) \textit{comparison of the performance of different components or subsystems within the same system}; and (d) \textit{comparison between measured and simulated PV system performance}. 

Those approaches might provide different perspectives over the same problem and potentially can complement each other. Nonetheless, the \textit{comparison between measured and simulated PV system performance} is very popular and has been explored by many authors in the literature \cite{davarifar_new_2013, garoudja_statistical_2017, jiang_automatic_2015, tadj_improving_2014, umana_detection_2016, garoudja_efficient_2016,gokmen_simple_2012,silvestre2013automatic, garoudja_statistical_2017, tadj_improving_2014}. This approach involves modeling the PV system behavior, calculating the deviation between measured and simulated outputs, and analyzing if the deviation is within an acceptable range. The simulation model is an especially attractive feature of this approach, it can also be used to provide the basis for accurate intraday or day ahead PV power forecasts or even long-term yield predictions.

Different methods can be used in each step of the process, and the selection of methods might influence the overall performance of the fault detection. For example, the modeling can be done using analytical, statistical or empirical models. The deviation in some publications is the difference between expected and measured output \cite{davarifar_new_2013, garoudja_statistical_2017, jiang_automatic_2015, tadj_improving_2014, umana_detection_2016}, while in others it is the ratio between measured and expected output \cite{garoudja_efficient_2016,gokmen_simple_2012,silvestre2013automatic, garoudja_statistical_2017, tadj_improving_2014}. The classification between normal and faulty behavior can be done using control charts or sophisticated machine learning algorithms.

It is unclear how the selection of these methods influences the performance of fault detection, especially due to the lack of a common basis for comparing the different methods, as well as a reference of actual faults to evaluate the effectiveness of fault detection, as pointed out by \cite{rapaport2021advanced}. 
This study works on this gap and seeks to provide insight into the influencing aspects in the design of AFD algorithms through practical experiments with data from the field.
A series of AFD algorithms have been designed by combining methods for modeling the PV system, calculating the deviation, and analyzing the deviation. The performance of each algorithm is tested using monitoring data from 80 commercially operated PV systems. The maintenance records are used to measure the algorithms' performance by quantifying the percentage of faults detected and the false alerts generated. 

This work is part of the research project OptOM (see acknowledgments), within which an integrated, cost-optimized O\&M concept for the economic lifetime of PV power plants will be developed. The study of fault detection complements previous studies on long-term performance loss \cite{kiefer2019degradation} in pursuing a cost-optimized maintenance plan to correct faults and reduce energy losses. See \cite{sarquis2021analysis} for previous publications on this topic within this project.

\section{Field data available} \label{sec:fielddata}

\begin{figure}[h]
    \centering
    \includegraphics[width=9cm]{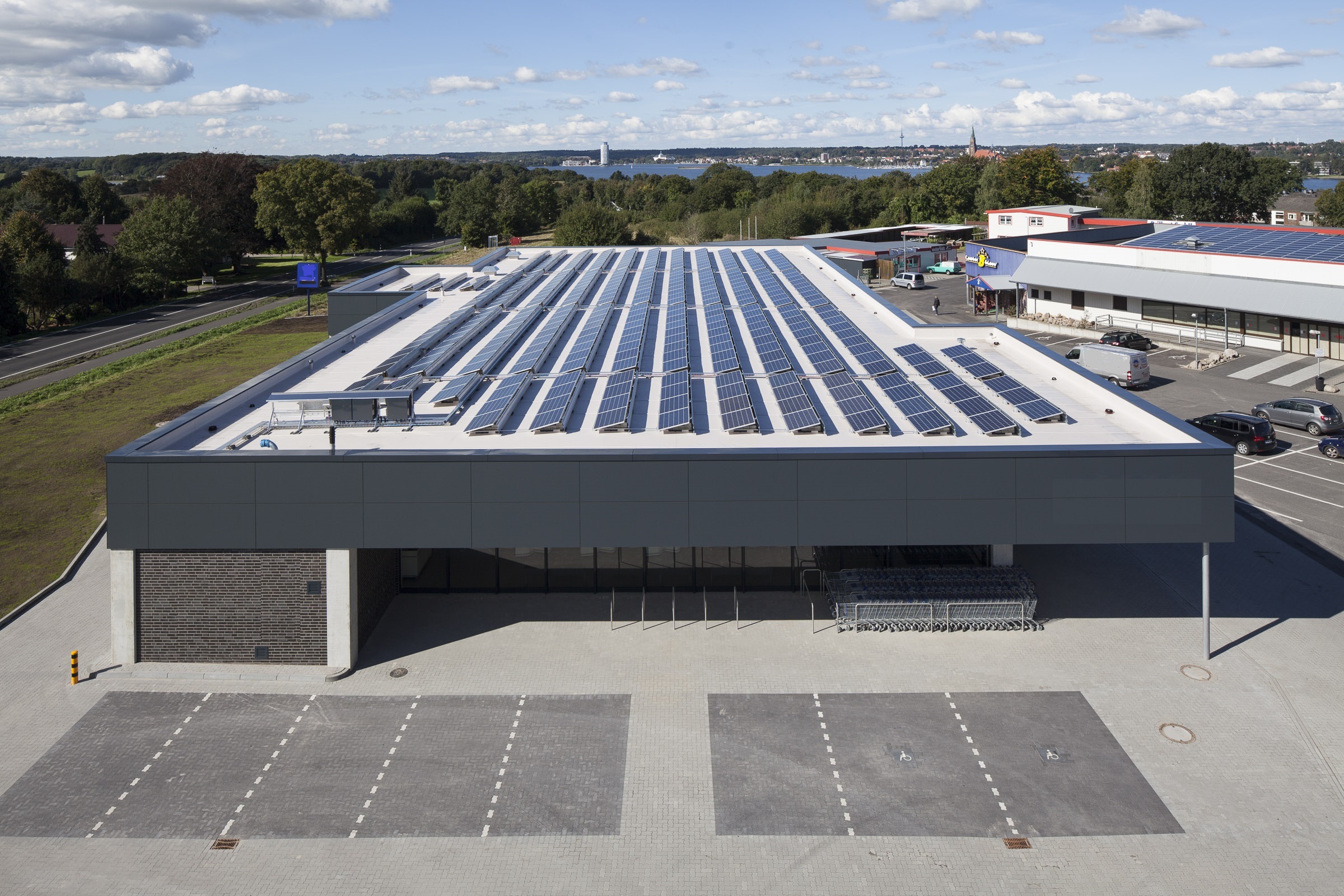}
    \caption{Example of PV system within the portfolio.}
    \label{fig:example_pvsystem}
\end{figure}

All field data in this study comes from 80 rooftop PV systems installed in Germany. One such system can be seen in Figure~\ref{fig:example_pvsystem}. The size of these systems ranges from 55 to 1550 kWp; 64\% have less than 300 kWp. All systems combined have 918 inverters and 135 420 PV modules. Most of the inverters (94\%) are string inverters rated up to 25 kWp, and the remaining 33 are central inverters with 400 to 600 kWp. The nominal power of PV modules ranges from 145 to 280 W, with the majority being 220 W.

The monitoring data used were collected from January 2016 to October 2020 (58 months). It comes mostly from the sensors available in the inverters, which measure the current, voltage, and power signals at the input and output of the device. Additionally, each system contains one or more irradiance sensors (reference cells) installed in the same plane as the PV panels. The data is acquired every minute or faster and averaged values are stored every 5 minutes. This monitoring setup resembles the common practices adopted in commercial medium-sized PV systems. 

The fault history of all 80 PV systems combined have 1529 unique monitoring and maintenance tickets, summing 17732 affected days from a total of 130832 recorded operating days. Tickets are created on the day the fault or suspicious behavior is identified and are repeated daily until the source of the problem is corrected. 

\subsection{Estimation of the energy losses} \label{sec:estimation_empirical}

\begin{figure}[h]
    \centering
    \includegraphics[width=9cm]{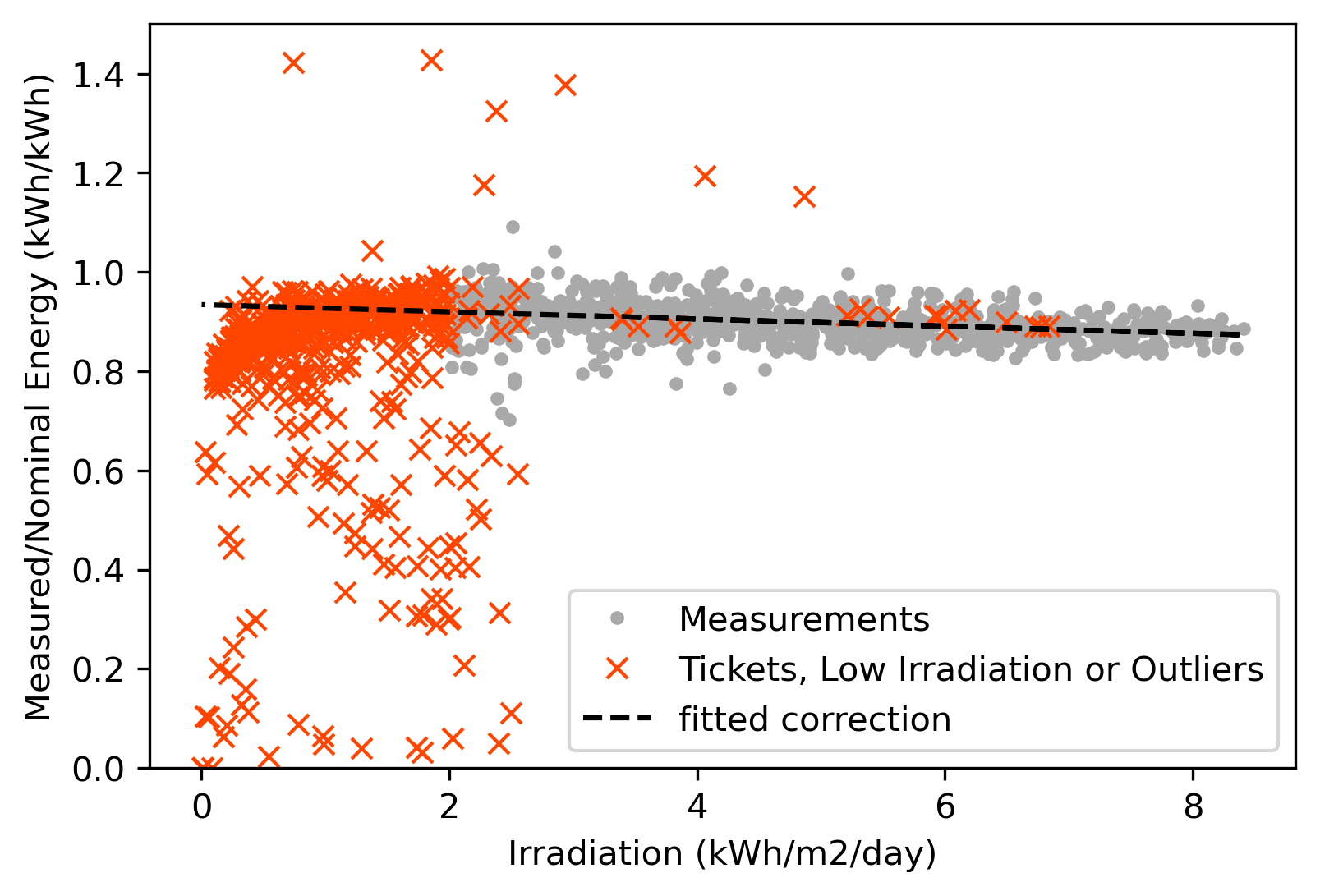}
    \caption{Modeling the correlation between measured and nominal energy ratio ($E_{meas}/E_{nom}$) and the irradiation $H_{POA}$. Days with tickets, low irradiance (bellow 2 kWh/m$^2$) or outlying values are ignored.}
    \label{fig:correction_factor}
\end{figure}

\begin{figure}[h]
    \centering
    \includegraphics[width=9cm]{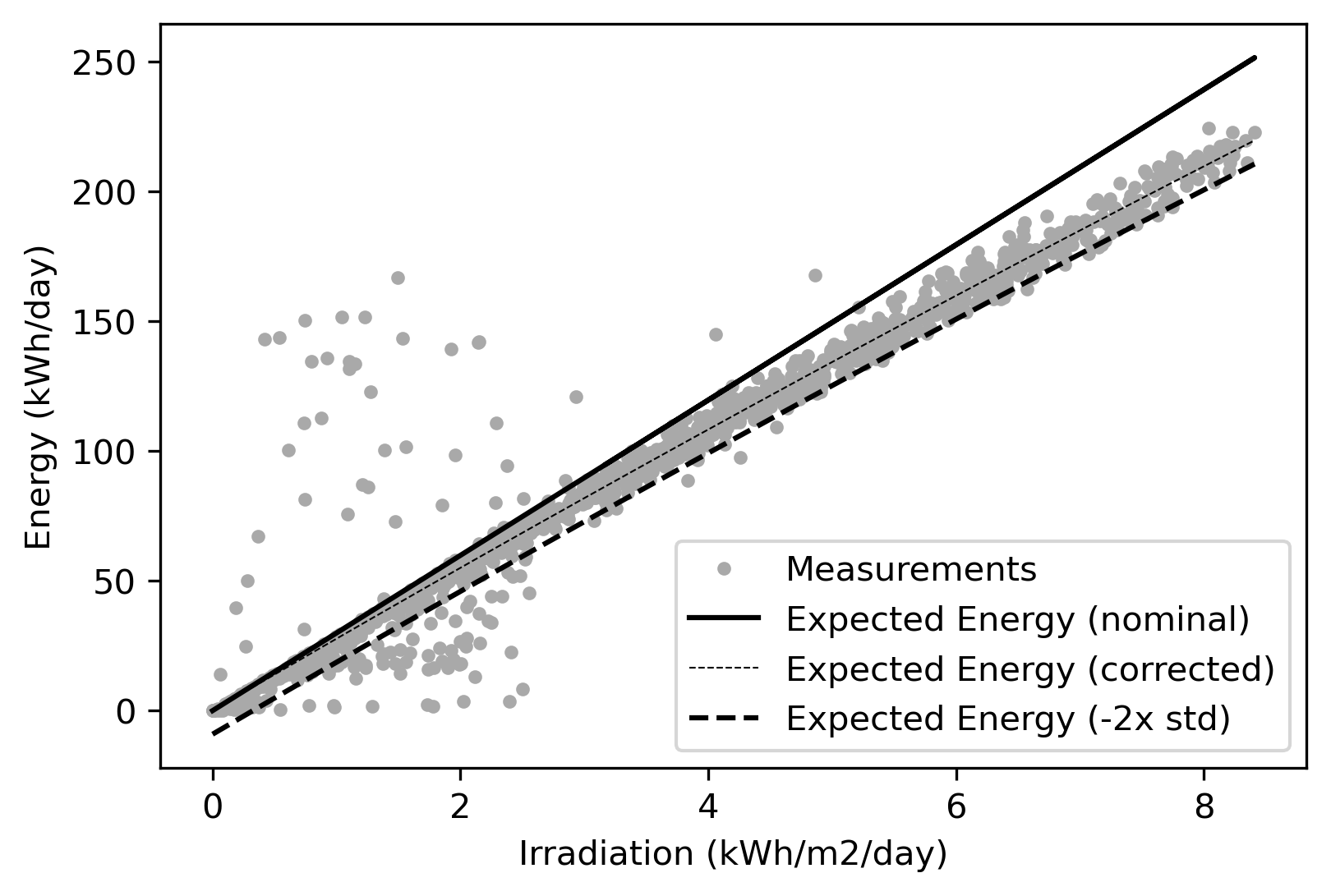}
    \caption{Comparison among the measured energy and expected energy estimation based on (a) nominal power, (b) corrected for irradiance and temperature and (c) discounted by two times the standard deviation.}
    \label{fig:exp_energy_corrected}
\end{figure}

For measuring the relevance of each daily ticket, a simple empirical method was applied to estimate the energy losses of the systems. The estimation starts with the calculation of the energy expected daily $E_{nom}$ according to the nominal power of the system ($P_{nom}$):

\begin{equation} \label{eq:nominal_exp_energy}
    E_{nom} = P_{nom} \times \frac{H_{POA}}{G_{STC}},
\end{equation}

where $H_{POA}$ is the daily in-plane irradiation available in kWh/m$^2$ and $G_{STC}$ is the irradiance at the standard test condition, 1 kW/m$^2$. The ratio between the measured energy $E_{meas}$ and $E_{nom}$ decreases with the $H_{POA}$, as can be seen in Figure~\ref{fig:correction_factor}. This seems reasonable, as with increased irradiation in general also ambient temperature and in particular the module temperature will be higher. 
This leads to reduced efficiency of the PV modules due to the negative temperature coefficient of power. Beside the temperature effect, other effects like inverter power limitation might contribute, too, but we estimate these effects to be of second order compared to the temperature effect. Effects like low-light efficiency of PV modules, cabling ohmic losses or reduced part-load efficiency of the inverters have the opposite influence and will contribute to the combined effect, too.
However, by modeling the combined effect through a linear regression and applying a correction factor $\theta$ our method implicitly considers the influence of temperature on the system behavior without having temperature data available. For the calculation of the coefficients $a$ and $b$ in (\ref{eq:correction}), data points from ticketed days, low irradiation days (bellow 2 kWh/m$^2$), or simply outliers, are ignored. The corrected expected energy for the day is then defined by (\ref{eq:exp_energy_corrected}). Figure~\ref{fig:exp_energy_corrected} depicts an example of the expected energy correction in comparison with the measurements.

\begin{equation} \label{eq:correction}
    \varphi(H_{POA}) = a \times H_{POA} + b
\end{equation}

\begin{equation} \label{eq:exp_energy_corrected}
    E_{exp} = E_{nom} \times \varphi(H_{POA})
\end{equation}

The positive difference between the expected and the measured daily energy can be interpreted as energy loss $E_{loss}$, as calculated in (\ref{eq:energy_loss_estimation}). To neglect small loss values within the uncertainty range, the expected values are reduced by two times the standard deviation $\sigma$, which is given by (\ref{eq:standard_deviation}). An example of nominal, corrected and reduced expected energy are depicted in Figure~\ref{fig:exp_energy_corrected}. The specific energy loss $SE_{loss}$ is then obtained by dividing $E_{loss}$ by the installed capacity of the system, as shown in (\ref{eq:sel_loss}).

\begin{equation} \label{eq:energy_loss_estimation}
    E_{loss} = E_{exp} - 2 \sigma - E_{meas}
\end{equation}

\begin{equation} \label{eq:standard_deviation}
    \sigma = \sqrt{\frac{1}{N} \sum_{i=1}^N (E_{exp,i} - E_{meas,i})^2}
\end{equation}

\begin{equation} \label{eq:sel_loss}
    SE_{loss} = E_{loss} / P_{nom}
\end{equation}

\begin{figure}[h]
    \centering
    \includegraphics[width=9cm]{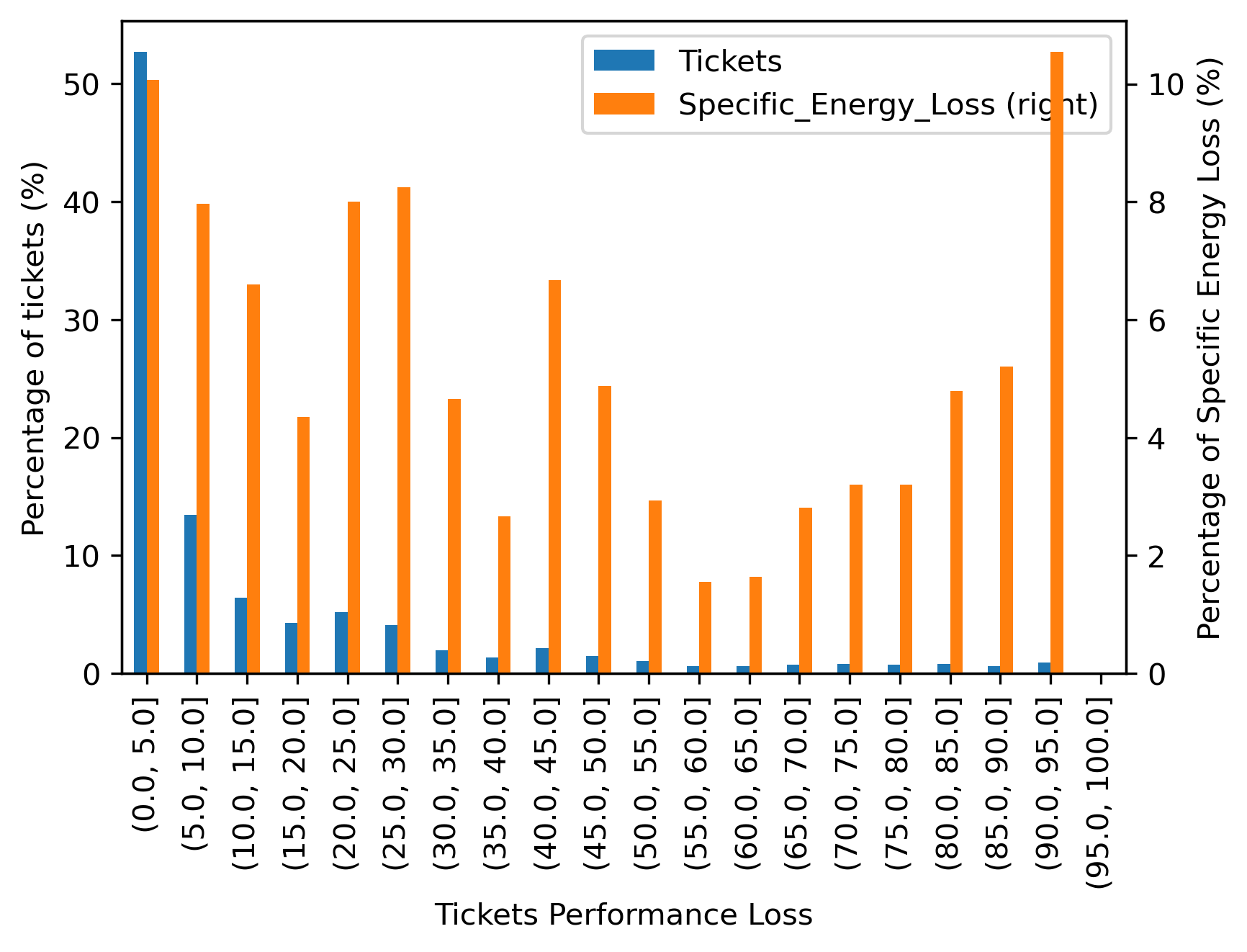}
    \caption{Distribution of tickets and specific energy loss according to tickets' performance loss.}
    \label{fig:sel_histogram}
\end{figure}

A ratio between the energy loss and the expected energy gives a sense of the daily performance loss $PL$ related to each ticket, as described in (\ref{eq:performace_loss}). Figure~\ref{fig:sel_histogram} depicts the distribution of tickets and specific energy loss according to performance loss intervals. It is observed that most of the tickets have low performance loss, being more than 50\% of the tickets limited to 5\% of performance loss. Nonetheless, the energy losses are spread across the entire range of performance losses. 

\begin{equation} \label{eq:performace_loss}
    PL = E_{loss} / E_{exp}
\end{equation}

\section{Automatic fault detection}

\begin{figure}[h]
    \centering
    \includegraphics[width=14cm]{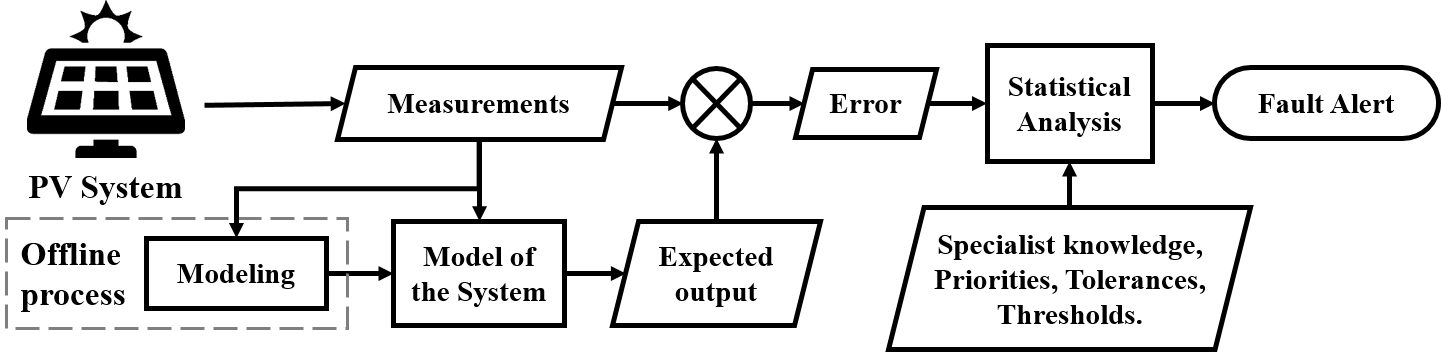}
    \caption{Block diagram of the fault detection process based on the comparison between measured and simulated PV system performance.}
    \label{fig:comparison_block_diagram}
\end{figure}

To detect faults automatically using the \textit{comparison between measured and simulated PV system performance}, first we use past electrical behavior and weather data to create a prediction model for the PV system. Then we use it to simulate the system output according to the weather data acquired subsequently. The deviation observed between the simulated and the measured output is then subjected to a statistical analysis that will trigger an alert in case abnormal deviation levels are found. The block diagram in Figure~\ref{fig:comparison_block_diagram} illustrated this process. 

The deviation can be calculated in absolute terms, as seen in \cite{davarifar_new_2013, garoudja_statistical_2017, jiang_automatic_2015, tadj_improving_2014, umana_detection_2016}, or in relative terms, like in \cite{garoudja_efficient_2016,gokmen_simple_2012,silvestre2013automatic, garoudja_statistical_2017, tadj_improving_2014}. Here we define the absolute deviation $Deviation_{abs}$ in (\ref{eq:absolute_deviation}) and the relative deviation $Deviation_{rel}$ in (\ref{eq:relative_deviation}). $Deviation_{abs}$ is normalized by the nominal power $P_{nom}$ to keep the deviation values within a similar range, regardless of the system size. The negative sign on both deviations should not be disregarded, so that the normal (or quasi-normal) distribution is preserved. Energy values must be quantified at regular time intervals to have consistent $Deviation_{abs}$ values for analysis.

\begin{equation} \label{eq:absolute_deviation}
    Deviation_{abs} = \frac{E_{meas} - E_{exp}}{P_{nom}}
\end{equation}

\begin{equation} \label{eq:relative_deviation}
    Deviation_{rel} = \frac{E_{meas} - E_{exp}}{P_{exp}} = \frac{E_{meas}}{E_{exp}} - 1
\end{equation}

In the following sections, we review different methods for modeling the PV system and for performing the statistical analysis of the deviation.

\subsection{Methods for modeling the PV system}
\label{sec:modelling}

The expected output of a PV system can be estimated using analytical modeling, i.e., calculating the output power based on equations describing the system’s physical characteristics. For example, the single or double diode model has been widely adopted to estimate the output characteristics of PV arrays. 
The parameters for these models can be calculated based on the information available in the PV module's datasheet \cite{davarifar_new_2013, umana_detection_2016, tadj_improving_2014} or estimated directly from the characteristic curves measured on the installed panels \cite{garoudja_statistical_2017, silvestre2013automatic}.
The simulation using PV module parameters given by the manufacturer does not consider the mismatch effects in the PV array configuration, resulting in an over-estimation of the output power \cite{chouder_automatic_2010}. On the other hand, it is rare to have IV curve data available for the installed PV array. For that reason, no analytical method was tested in this study.

Another commonly used approach adopts statistical modeling to learn the behavior of the system based on its measurements. The minimum information needed is the irradiance and the targeted output variable, typically AC power. When available, PV module temperature measurements can refine the modeling accuracy. 
Within the reviewed literature, this approach has been implemented using different algorithms, but recent studies have shown that there is no significant difference in the way each algorithm extracts information from the input data, resulting in very similar performance \cite{rapaport2021advanced}. 
Considering this, for this study, we selected a type of linear regression, a second-order polynomial regression model, because it is simple to implement and define coefficients. The output power is calculated based on the irradiance according to the expression (\ref{eq:polyreg_model}).

\begin{equation} \label{eq:polyreg_model}
    P(t) = a_0 + a_1 G(t) + a_2 G(t)^2,
\end{equation}

In \cite{lazzaretti_monitoring_2020}, a linear auto-regressive model with exogenous inputs (ARX) is used to estimate the AC power output using the latest AC power and irradiance measurements. The ARX model estimates the current output power ($P(t)$) based on the previous measurements of output power ($P(t-1)$ and $P(t-2)$) and also irradiance ($G(t)$ and $G(t-1)$), as described by (\ref{eq:arx_model}). This modeling method also has been considered for the tests.

\begin{equation} \label{eq:arx_model}
    P(t) = a_1 P(t-1) + a_2 P(t-2) + b_0 G(t) + b_1 G(t-1),
\end{equation}

An empirical approach to estimate the output power of a PV system was already introduced in section \ref{sec:estimation_empirical}. It combines analytical equations from the PV domain with learning methods to adjust their coefficients according to measurements. The empirical approach is explored in \cite{heydenreich2008describing}, where the authors propose two equations based on three adjustable parameters to model the efficiency of the PV modules according to irradiance and PV module temperature, and then estimate the power output according to the module area installed. In a more sophisticated way, in \cite{guzman2020genetic} a chain of physical models is used to estimate several energy conversion steps – from irradiance to output AC power. The models’ parameters are identified using an optimization procedure applying a genetic algorithm (GA). Both methods require PV module temperature measurements, which are not available in this study and therefore could not be tested. In the experiments, the empirical modeling approach is tested using the method described in section \ref{sec:estimation_empirical}.

\subsection{Statistical analysis} \label{sec:statistical_analysis}

The statistical analysis is the step in which the deviation values are analyzed and classified between fault and normal behavior. The decision boundaries must be defined considering the variable nature of the weather, which is reflected in the PV systems’ behavior. Historically other industries than PV have developed tools to tackle similar problems. For example, statistical process control (SPC) tools have been developed to detect the occurrence of process shifts so that investigation and corrective action may be undertaken. Those tools are based on sound statistical principles and can be applied to any process \cite{montgomery2020introduction}. One of the main SPC tools is the control chart, and one example of its application for fault detection in PV systems is seen in \cite{garoudja_statistical_2017}.

As an alternative to control charts, today sophisticated data analyses and decision-making can be done using machine learning algorithms. Clustering, for example, is a specific type of machine learning algorithm capable of segmenting data into mutually exclusive homogeneous groups (clusters) according to their characteristics, which can be used to distinguish faults from normal behavior \cite{rapaport2021advanced}. It does not require labeled training data or prior knowledge of the fault behavior to perform the classification.

In the following sections, more details are given on how control charts and clustering algorithms can be used in the design of AFD algorithms.

\subsubsection{Control charts}

\begin{figure}[h]
    \centering
    \includegraphics[width=9cm]{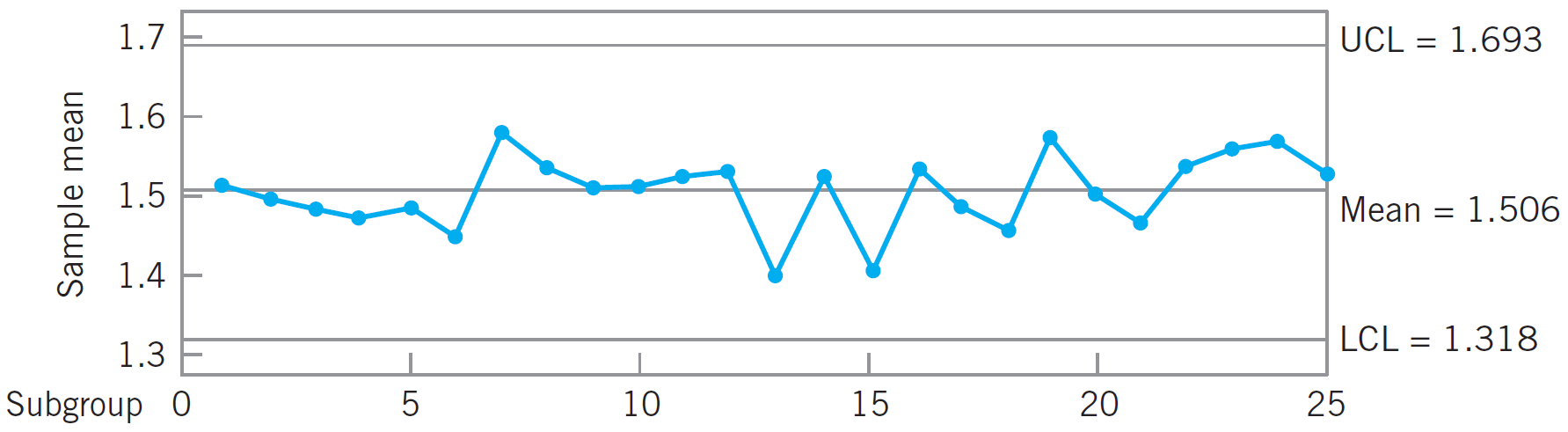}
    \caption{Example of control chart \cite{montgomery2020introduction}.}
    \label{fig:example_controlchart}
\end{figure}

The control chart is an online process-monitoring technique widely used to quickly detect unacceptable disturbances in the process performance that could be associated with errors, faults, or external influences. Figure~\ref{fig:example_controlchart} shows an example of a control chart. The centerline represents the average $\mu$ of the process monitored, and the two control limits -- upper (UCL) and lower (LCL) -- are chosen so that nearly all of the sample points will fall between them if the process is in control. The control limits are expressed in standard deviation $\sigma$ units, as described in (\ref{eq:shewhart}). The term $L$ defines the width of the tolerance window and $n$ is the sample size. 
Control charts developed according to these principles are often called \textit{Shewhart control charts} \cite{montgomery2020introduction}.

\begin{equation} \label{eq:shewhart}
    \begin{split}
        UCL = & \mu + L \frac{\sigma}{\sqrt{n}} \\
        Center \, line = & \mu \\
        LCL = & \mu - L \frac{\sigma}{\sqrt{n}}
    \end{split}
\end{equation}

\begin{figure}[h]
    \centering
    \includegraphics[width=9cm]{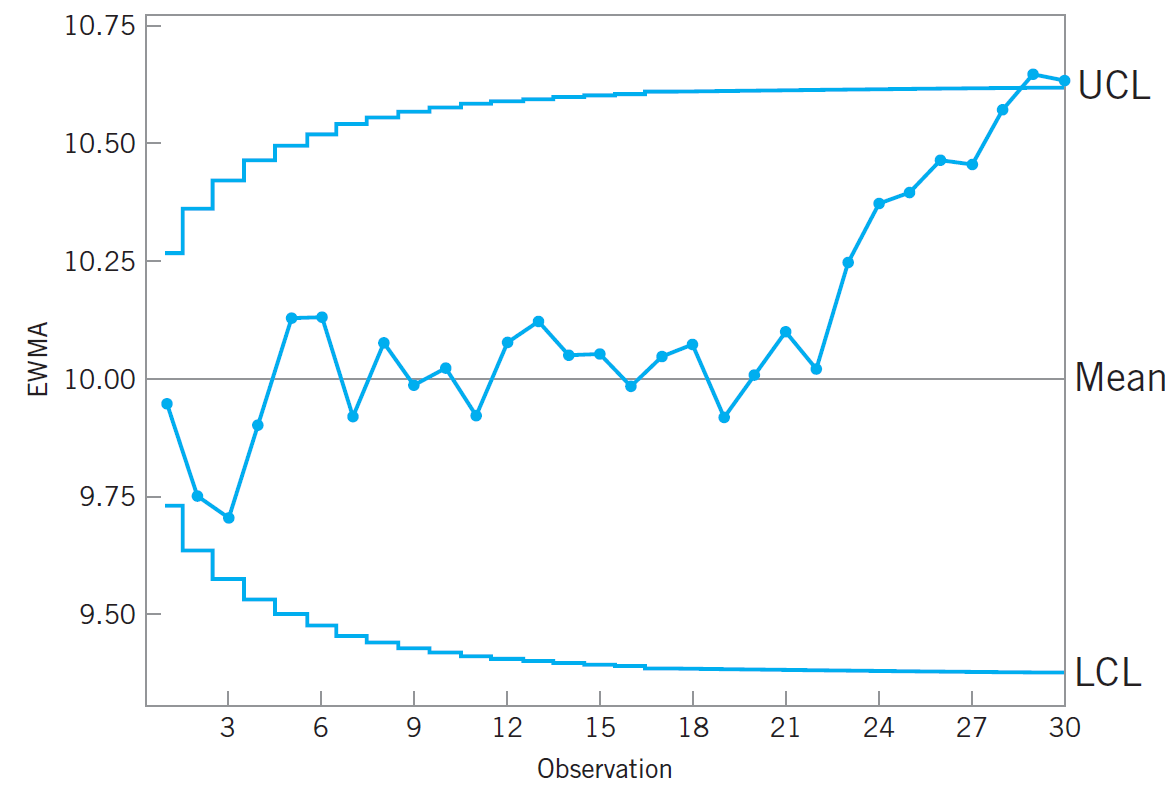}
    \caption{Example of an EWMA control chart \cite{montgomery2020introduction}.}
    \label{fig:example_ewmacontrolchart}
\end{figure}

Shewhart control charts are easy to construct and interpret, but they are not very sensitive to small process shifts. The \textit{exponentially weighted moving average} (EWMA) control charts are more effective in those cases \cite{montgomery2020introduction,garoudja_statistical_2017}. 
The centerline and control limits of this chart are defined in (\ref{eq:ewma_chart}). They monitor the EWMA $z_t$ of the monitored process $x_t$, as defined in (\ref{eq:ewma_definition}). The initial value $z_0$ is usually set to be the process average $\mu$. The parameter $\lambda \in [0; 1]$ determines how fast EWMA "forgets" the previous observations.
The upper and lower control limits (UCL/LCL) are defined according to the mean $\mu$, a tolerance $L$ and the standard deviation $\sigma_{t}$ for a time $t$, which is defined in (\ref{eq:ewma_std}) according to the process standard deviation $\sigma_{0}$ and $\lambda$. The last term in equation (\ref{eq:ewma_std}) approaches unity as $t$ gets larger, after the EWMA control chart has been running for some time, the control limits will approach steady-state values \cite{montgomery2020introduction}. See Figure~\ref{fig:example_ewmacontrolchart}.

\begin{equation} \label{eq:ewma_chart}
    \begin{split}
        UCL = & \mu + L \sigma_{t}\\
        Center \, line = & \mu\\
        LCL = & \mu - L \sigma_{t}
    \end{split}
\end{equation}

\begin{align} \label{eq:ewma_definition}
    \begin{cases}
        z_t = \lambda x_t + (1-\lambda) z_{t-1} & \text{ if } t > 0; \\ 
        z_0 = \mu & \text{ if } t = 0. 
    \end{cases}
\end{align}

\begin{equation} \label{eq:ewma_std}
    \sigma_{t} = \sigma_0 \sqrt{\frac{\lambda}{(2-\lambda)}[1-(1-\lambda)^{2t}]}
\end{equation}

The control limits must be adjusted to detect deviations from normal behavior and, at the same time, reduce the probability of \textit{false alarms} and \textit{false negatives}.
Increasing $L$ and moving the control limits farther from the center line decreases the risk of raising a false alarm, but also increases the chances of missing a fault occurrence, i.e., having a false negative. 
A widespread practice is to assume that the monitored signal is normally distributed and adopt three-sigma control limits.

Some economic aspects should also be considered when defining the control limits. In the PV monitoring domain, it can be expensive to mobilize a field inspection to investigate a false alarm. Therefore it might be reasonable to widen the control limits beyond the three-sigma, say 3.5-sigma. The downside is that this can cause some delay in alerts or even the inability to detect subtle system performance losses.

The control charts can be built to monitor every single measurement individually (sample size $n=1$) or in \textit{rational subgroups} (sample size $n>1$). As a good practice, measurements are grouped according to their proximity in time, say within a sampling interval. Each sample should represent the process natural variability, thus the sampling interval should not be too long, so that it does not contain drifts in the system's behavior \cite{montgomery2020introduction}. 
When using rational subgroups, the values monitored in the charts are the average of the observations in each sample. 

For the design of control charts, the next section describes how to estimate $\mu$ and $\sigma$ according to the fundamentals of statistical process control.

\subsubsection{Estimation of the mean and standard deviation} \label{sec:estimation_mu_sigma}

The statistical control of a process is based on its mean $\mu$ and standard deviation $\sigma$. These values are usually unknown, therefore they can be estimated from mean $\bar{x}$ and standard deviation $s$ of previous samples using equations (\ref{eq:mean_estimator}) and (\ref{eq:sigma_estimator}), respectively, as described in \cite{montgomery2020introduction}.

\begin{equation} \label{eq:mean_estimator}
    \hat{\mu} = \bar{\bar{x}}
\end{equation}

\begin{equation} \label{eq:sigma_estimator}
    \hat{\sigma} = \frac{\bar{s}}{c}
\end{equation}

In (\ref{eq:sigma_estimator}), $c$ is a constant that depends on the sample size $n$ according to:

\begin{equation}
    c = \left (  \frac{2}{n-1}  \right )^{1/2} \frac{\Gamma (n/2)}{\Gamma[(n-1)/2]} \sigma.
\end{equation}

As described in \cite{montgomery2020introduction}, for large sample sizes ($n > 25$), the value of $c$ can be approximated by 

\begin{equation}
    c \cong \frac{4(n-1)}{4n-3}.
\end{equation}

In (\ref{eq:mean_estimator}), $\bar{\bar{x}}$ is the grand average of the samples, i.e., the average of the samples' average, say:

\begin{equation} \label{eq:samples_grand_average}
    \bar{\bar{x}} = \frac{\bar{x}_1 + \bar{x}_2 + ... + \bar{x}_m}{m},
\end{equation}

where $m$ is the number of samples available.

In (\ref{eq:sigma_estimator}), $\bar{s}$ is the average of the samples' standard deviation:

\begin{equation}
    \bar{s} = \frac{s_1 + s_2 + ... + s_m}{m}
\end{equation}

The estimation of $\bar{s}$ using the usual quadratics estimator applied over all preliminary data is prone to error.
If any change in the mean deviation has occurred, this will cause $\bar{s}$ to be overestimated. Estimating the standard deviation of samples separately ensures the control limits will be defined based only on the within-sample variability, which better reflects the process natural variability. Moreover, these estimates should usually be based on at least 20 to 25 samples. 

The range $R$ of a sample of size $n$ is given by 

\begin{equation} \label{eq:range_definition}
    R = \textup{max}_{t=1}^n x_t - \textup{min}_{t=1}^n x_t
\end{equation}

As described in \cite{montgomery2020introduction}, for sample sizes less than or equal to six ($n \le 6$), an unbiased estimation of the standard deviation $\hat{\sigma}$ can be obtained from 

\begin{equation} \label{eq:R2sigma}
    \hat{\sigma} = \frac{\bar{R}}{d},
\end{equation}

where $d$ is a constant that depends on the sample size and $\bar{R}$ is the average range given by

\begin{equation}
    \bar{R} = \frac{R_1 + R_2 + ... + R_m}{m}.
\end{equation}

The values of $d$ for various sample sizes are given in Table~\ref{tab:d_values}. 

\begin{table}[]
\centering
\caption{Values of $d$ according to the sample size $n$. Extracted from \cite{montgomery2020introduction}}
\small
\label{tab:d_values}
\begin{tabular}{ll}
\hline
$n$ & $d$ \\ \hline
2   & 1.128  \\
3   & 1.693  \\
4   & 2.059  \\
5   & 2.326  \\
6   & 2.534  \\ \hline
\end{tabular}
\end{table}

Suppose the control charts should be designed to monitor individual measurements. Then the process standard deviation $\sigma$ can be estimated using (\ref{eq:R2sigma}) being the range the difference between two adjacent measurements, as described in (\ref{eq:R_for_adjacent}). The value of $d$, in this case, should be 1.128. The control charts limits are then defined for sample size of one ($n=1$).

\begin{equation} \label{eq:R_for_adjacent}
    R = x_t - x_{t-1}
\end{equation}

\subsubsection{Clustering algorithm}

In this study, we analyze the deviation data using \textit{k-means}, a clustering algorithm that segregate the observations into a predefined number $k$ of clusters while minimizing within-cluster variances via squared Euclidean distances \cite{wiki_kmeans_2021}. 

\begin{figure}[h]
    \centering
    \includegraphics[width=14cm]{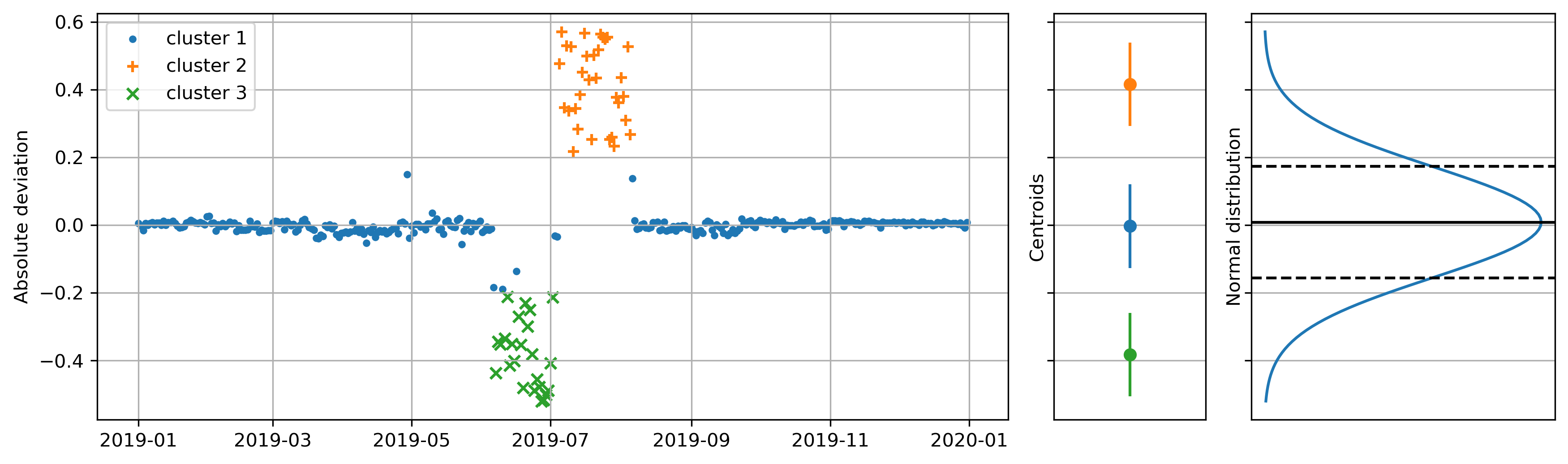}
    \caption{Example of deviation values classified in three clusters (k-means with $k=3$), their centroids and the normal distribution associated with the average and standard deviation.}
    \label{fig:clustering_k3}
\end{figure}

The cluster containing the deviation values associated with normal behavior is expected to have its centroid close to the process average $\mu$ (estimated according to section \ref{sec:estimation_mu_sigma}). The faulty points would be located above or below the process mean, forming two other clusters, thus the \textit{k-means} algorithm is initialized with $k=3$ clusters. Figure~\ref{fig:clustering_k3} depicts a classification where cluster 1 represents the normal behavior.

\begin{figure}[h]
    \centering
    \includegraphics[width=14cm]{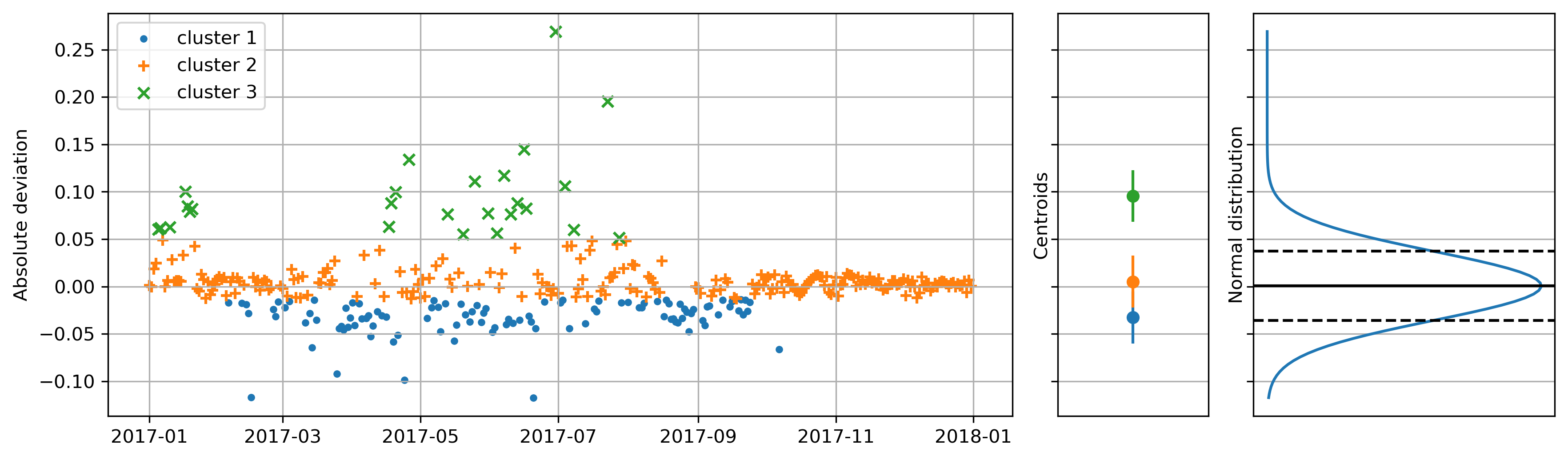}
    \caption{Example of deviation values classified into three clusters (k-means with k=3) resulting in two very close centroids.}
    \label{fig:clustering_k3_fail}
\end{figure}

\begin{figure}[h]
    \centering
    \includegraphics[width=14cm]{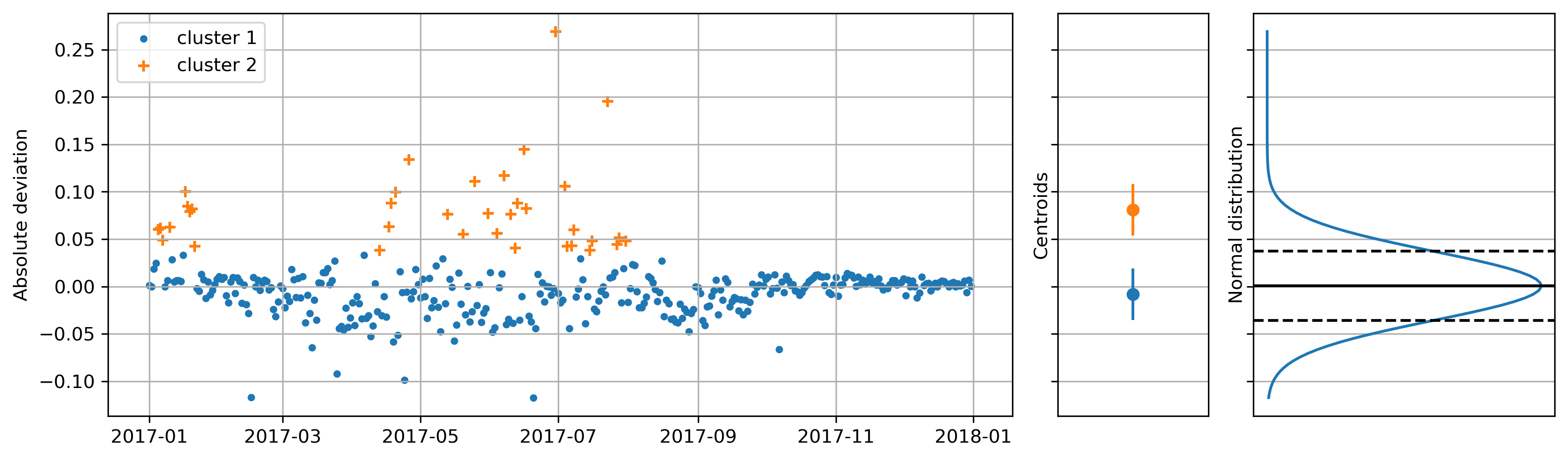}
    \caption{Example of deviation values reclassified into two clusters (k-means with k=2) resulting in properly separated centroids.}
    \label{fig:clustering_k2}
\end{figure}

In case the classification results in very close centroids (below a minimum distance), as shown in Figure~\ref{fig:clustering_k3_fail}, the number of clusters should be reduced to get a better classification of the data points. Figure~\ref{fig:clustering_k2} shows the reclassification into two groups with properly separated centroids. 

\begin{figure}[h]
    \centering
    \includegraphics[width=6cm]{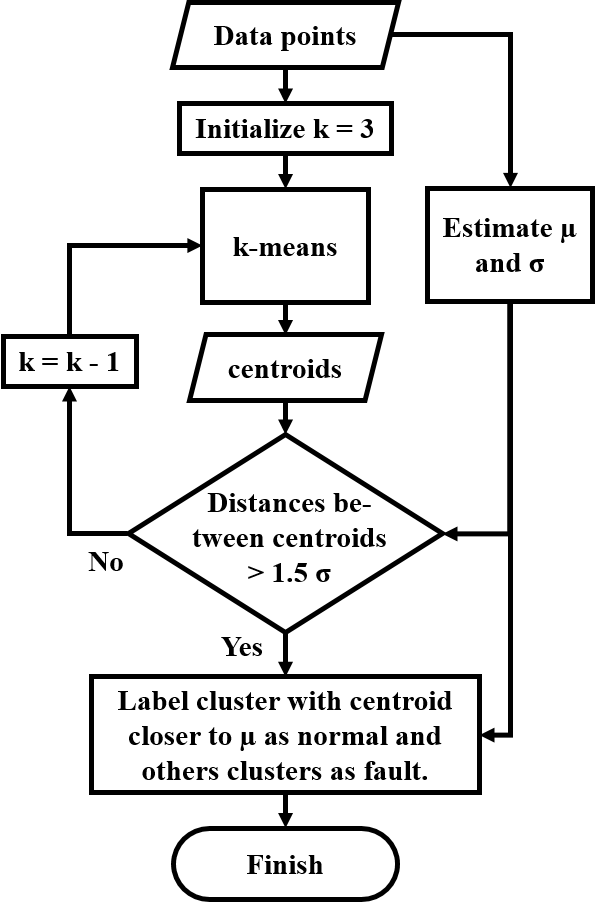}
    \caption{Procedure for fault detection using clustering algorithm.}
    \label{fig:kmeans_flowchart}
\end{figure}

The minimum distance is an input parameter that needs to be fine-tuned. We considered it to be $1.5 \times \sigma$, as estimated according to section \ref{sec:estimation_mu_sigma}. Figure~\ref{fig:kmeans_flowchart} describes this procedure in a flowchart.

\section{Experiments methodology}

\subsection{Methods tested}

A complete AFD algorithm combines a method for estimating the output and a method for perform the statistical analysis. 
This study combines the methods previously described and their variations to create a series of AFD algorithms.
Using the available data described in Section~\ref{sec:fielddata}, each AFD algorithms is implemented and tested.

The modeling methods considered for estimating of expected output are (see section \ref{sec:modelling} for details):
\begin{itemize}
    \item Auto-regression with exogenous inputs (ARX)
    \item Polynomial regression (PolyReg)
    \item Empirical
\end{itemize}

For measuring the level of accuracy provided by the different modeling methods, we calculate the mean absolute percentage deviation (MAPD) between the measured $P_{i}$ and simulated $\hat{P_{i}}$ output measurements. These metrics are calculated using equation (\ref{eq:mape_formula}).

\begin{equation} \label{eq:mape_formula}
    MAPD =  \frac{100}{n}\sum_{i=1}^{n}\frac{\left|P_{i} - \hat{P_{i}}\right|}{\left|P_{i}\right|}
\end{equation}

The methods considered for the statistical analysis are (see section \ref{sec:statistical_analysis} for details):
\begin{itemize}
    \item Shewhart chart
    \item EWMA chart
    \item k-means clustering
\end{itemize}

The control limits in the control charts adopts $L=3.5 \sigma$. For the clustering algorithm, the minimum distance between centroids is $1.5 \sigma$.

The deviation between simulated and measured output was considered in two ways: \textit{absolute deviation}, as per (\ref{eq:absolute_deviation}), and \textit{relative deviation}, as per (\ref{eq:relative_deviation}).
Alternatively, we tested  the performance ratio $PR$ as input for the statistical analysis. The $PR$ calculation is described in (\ref{eq:PR_definition}) as defined in IEC 61724 \cite{iec2021photovoltaic}.

\begin{equation} \label{eq:PR_definition}
    PR = \frac{E}{P_{nom}} \frac{G_{STC}}{H_{POA}},
\end{equation}

where $E$ is the PV system output energy.

For the statistical analysis, we have considered four ways of grouping the measurements:
\begin{itemize}
    \item \textit{5 min single}: Each measurement is analyzed individually, i.e. sample size $n$ is equal to one;
    \item \textit{30 min group}: The measurements are grouped in small rational subgroups every 30 minutes (sample size $n$ is six), and the process standard deviation is estimated using the samples' range;
    \item \textit{daily group}: The measurements are grouped per day, the sample size is variable but always larger than 25, thus the process standard deviation is estimated using the samples' standard deviation;
    \item \textit{daily single}: The measurements are aggregated in daily values, the deviation is calculated per day and analyzed individually, i.e. sample size $n$ is equal to one;
\end{itemize}

\subsection{Data quality check and filtering}

All measurements used for implementing and testing the AFD algorithms passed through a quality check procedure that performs the list-wise exclusion of data points which values falls outside the feasible physical limits, as recommended by \cite{livera2021data}. 
We also excluded data points acquired during night time and data points where the irradiance measured is below 50 W/m$^2$.
In applications that demand high quality standards, more elaborated physical checks should be adopted, e.g. the 'QCPV' quality control algorithm \cite{killinger2017qcpv}.

In the analysis of each system, the first 365 days of measurements were used to train the predictive model and to estimate the process mean and standard deviation for the statistical analysis.

When calculating the relative deviation, we filter out the data points for which the expected power is less than 5\% of the nominal power in order to avoid small denominators and ensure numerical stability. 

\subsection{Evaluation of the fault detection performance}

\begin{figure}[h]
    \centering
    \includegraphics[width=7cm]{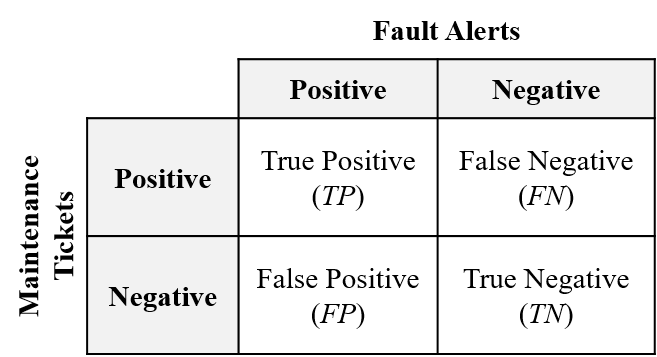}
    \caption{Confusion matrix.}
    \label{fig:conf_matrix}
\end{figure}

The performance of each AFD algorithm is evaluated by comparing its alerts with the maintenance tickets in all 80 PV systems for the whole period of five years. The maintenance tickets are written daily, thus the fault alerts from the AFD algorithms are also given on a daily basis. Each daily fault alert is classified as True Positive ($TP$), False Positive ($FP$), True Negative ($TN$), or False Negative ($FN$) according to the confusion matrix logic, as depicted in Figure~\ref{fig:conf_matrix}. The $TP$ are the detected faults, and $TN$ are normal operations. The $FN$ are the undetected faults, and $FP$ are the false alarms.

The True Positive Rate ($TPR$), also known as Sensitivity, and the True Negative Rate ($TNR$), also known as Specificity, are defined in (\ref{eq:sen_definition}) and (\ref{eq:spe_definition}), respectively. The sensitivity measures how many of the existing faults were correctly detected. The specificity measures how many of the negatives are indeed normal behavior.

\begin{equation} \label{eq:sen_definition}
    \textrm{Sensitivity} = \frac{TP}{TP + FN}
\end{equation}

\begin{equation} \label{eq:spe_definition}
    \textrm{Specificity} = \frac{TN}{TN + FP}
\end{equation}

Not all tickets in the maintenance records are associated with problems that significantly affect system performance.
The daily specific energy loss ($SE_{loss}$) estimated in section \ref{sec:estimation_empirical} indicates the relevance of the maintenance ticket in terms of system performance. A sensitivity weighted by the ticket relevance can be calculated using (\ref{eq:weighted_sen_definition}), where $TP_{rel}$ is the sum of the relevance of all true positive alerts, and $FN_{rel}$ is the sum of the relevance of all false negative alerts.

\begin{equation} \label{eq:weighted_sen_definition}
    \textrm{Weighted \, Sensitivity} = \frac{TP_{rel}}{TP_{rel} + FN_{rel}}
\end{equation}

\begin{figure}[h]
    \centering
    \includegraphics[width=7cm]{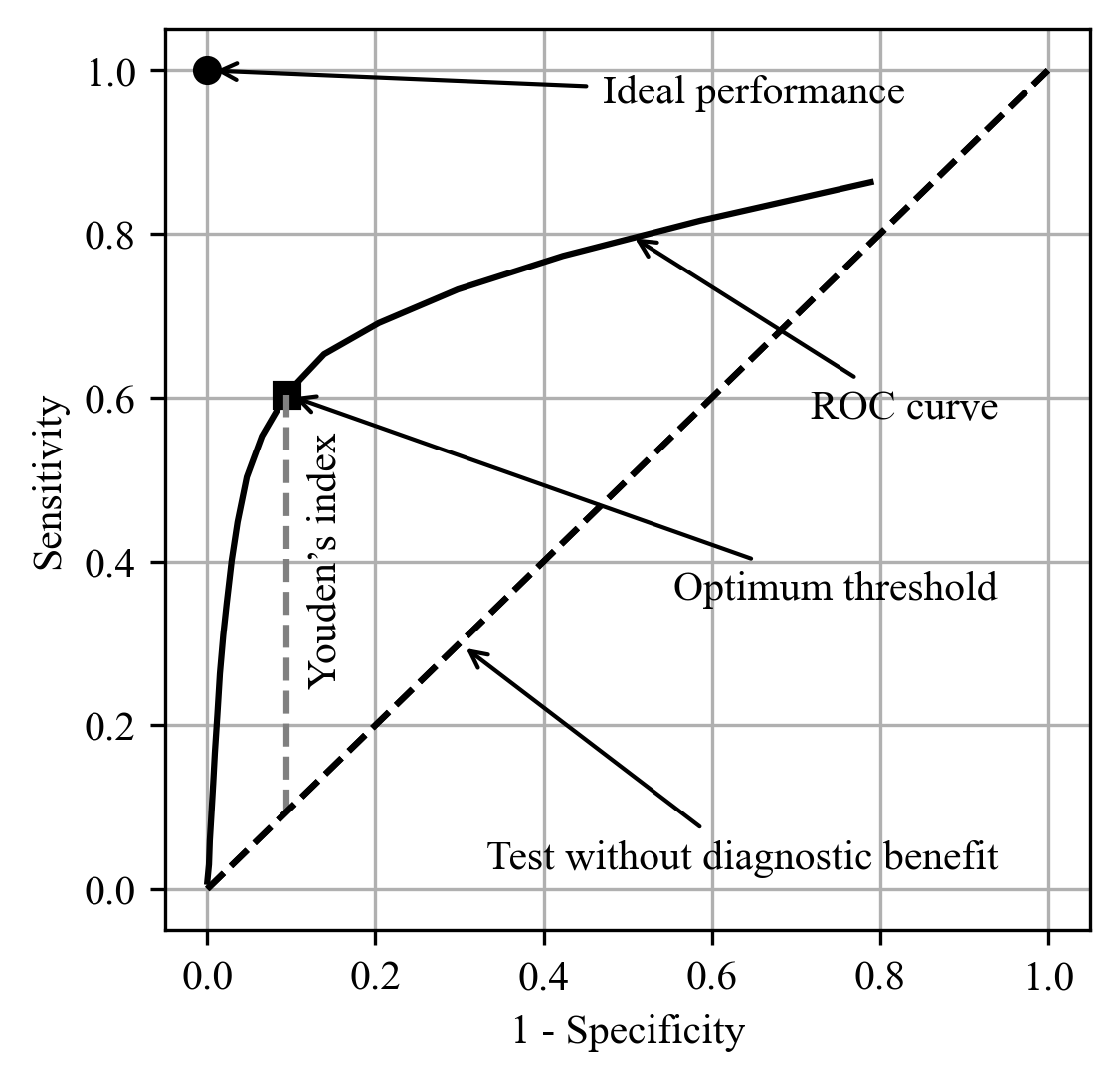}
    \caption{Example ROC curve highlighting the optimal threshold according to the Youden's index.}
    \label{fig:ROC_fundamentals}
\end{figure}

The performance of the AFD algorithms tested are evaluated using a Cartesian plane with the Sensitivity in the \textit{y} axis and the False Positive Rate ($FPR$), which is equivalent to $1 - \textrm{Specificity}$, in the \textit{x} axis.
The ideal performance is located at the top left corner of this graph, as depicted in Figure~\ref{fig:ROC_fundamentals}.

For the cases where the measurements are grouped in \textit{5 min single} or in \textit{30 min group}, the statistical analysis results are given in percentage of faulty samples observed in a day. These values are translated into Boolean daily fault alerts by applying a threshold. The performance of the AFD algorithm for all possible thresholds can be visualized as a curve in the Cartesian plane, the receiver-operator characteristic (ROC) curve. See Figure~\ref{fig:ROC_fundamentals}. The ROC was originally developed for operators of military radar receivers to measure the performance of binary classifiers \cite{fawcett2006introduction}.

From the ROC curve, we can find the optimum threshold that gives the best balance between sensitivity and specificity.
One criteria commonly adopted to find the optimum threshold is to maximize the Youden's $J$ statistic, also known as the Youden's index \cite{youden1950index}. Its definition and the threshold optimization criteria are described in (\ref{eq:youden_index}) and (\ref{eq:optimum_threshold}), respectively.
Graphically, the Youden's index is equivalent to the vertical distance between the ROC curve and the diagonal, as depicted in Figure~\ref{fig:ROC_fundamentals}. The largest Youden's index indicates the point on the ROC curve associated with the optimal limit.

\begin{equation} \label{eq:youden_index}
    \begin{split}
        J & = \textrm{Sensitivity} + \textrm{Specificity} - 1 \\ 
          & = TPR - FPR
    \end{split}
\end{equation}

\begin{equation} \label{eq:optimum_threshold}
    threshold_{optimum} = \max_{0 \leq threshold \leq 1} J
\end{equation}

\section{Results and discussion}

\subsection{Modeling accuracy}

\begin{figure}[h!]
    \centering
    \includegraphics[width=4.5cm]{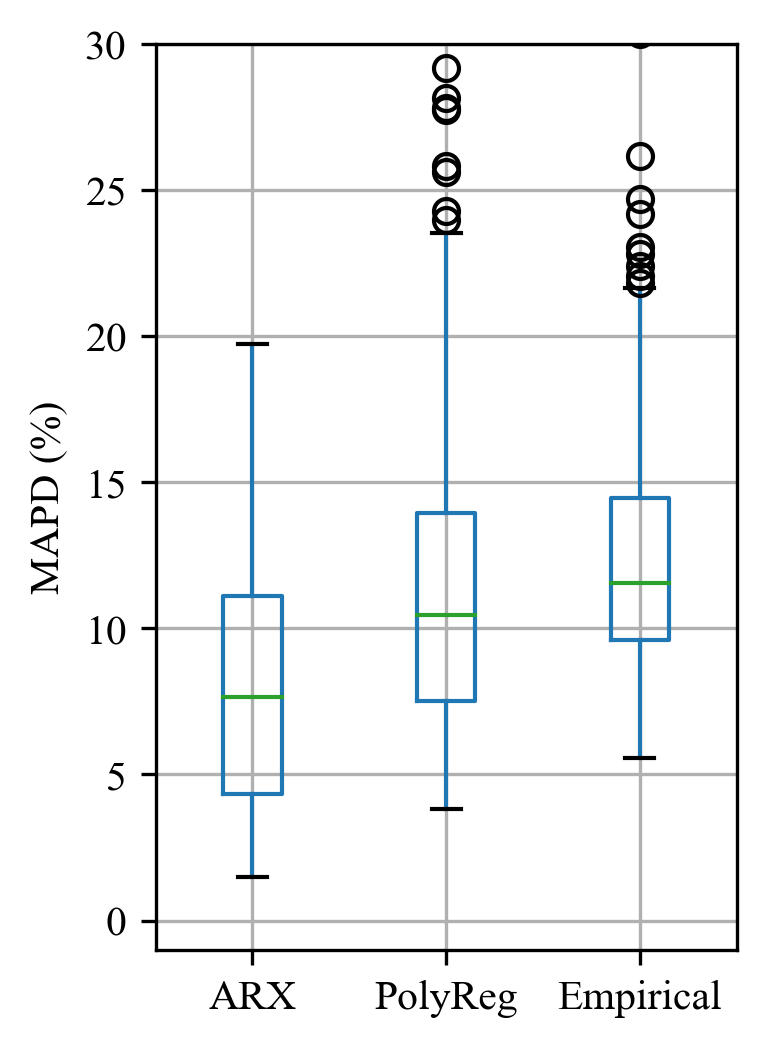}
    \caption{Expected output estimation accuracy according to the modeling tool.}
    \label{fig:exp_accuracy}
\end{figure}

First, we analyze the modeling accuracy according to the method used. The $MAPD$ ranges obtained for each modeling method are shown in Figure~\ref{fig:exp_accuracy}. The ARX model had the lowest deviations (median of 7.64\%). The other methods resulted in moderate deviations (median of 10.5\% and 11.6\%). In comparison with other PV simulation tools, those methods offer medium to low accuracy. In the authors' analysis, this is mainly due to the lack of temperature measurements and physical modeling. For example, a method based on a chain of physical models mentioned in Section~\ref{sec:modelling} is said to achieve $MAPD$ lower than 6\% \cite{guzman2020genetic}. 

\subsection{Statistical analysis using control charts}

\begin{figure}[hp]
    \centering
    \includegraphics[width=13cm]{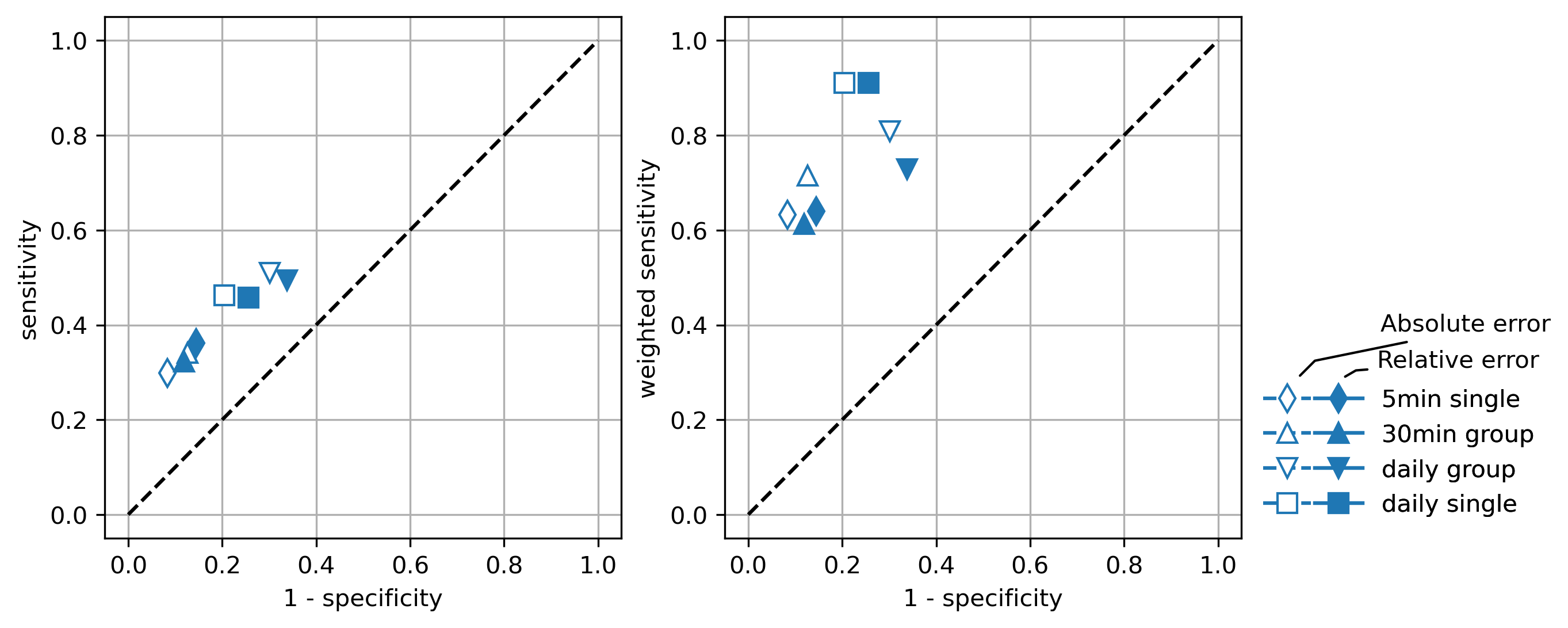}
    \caption{Sensitivity and specificity for all AFD algorithms adopting Shewhart charts and ARX modeling method.}
    \label{fig:ROC_shewhart_ARX_all}
\end{figure}

The performance of AFD algorithms that use control charts for statistical analysis depends on the type of chart, how the samples are grouped and the type of deviation analyzed. 
Figure~\ref{fig:ROC_shewhart_ARX_all} depicts the performance observed for the AFD algorithms based on \textit{Shewhart} charts with all four types of grouping (\textit{5 min single}, \textit{30 min group}, \textit{daily group}, and \textit{daily single}) and both types of deviations (\textit{absolute} and \textit{relative}).
All these algorithms adopt the same modeling method (the most accurate one, ARX model). The influence of the modeling accuracy will be discussed in Section~\ref{sec:impact_modeling}.

\begin{figure}[hp]
    \centering
    \includegraphics[width=13cm]{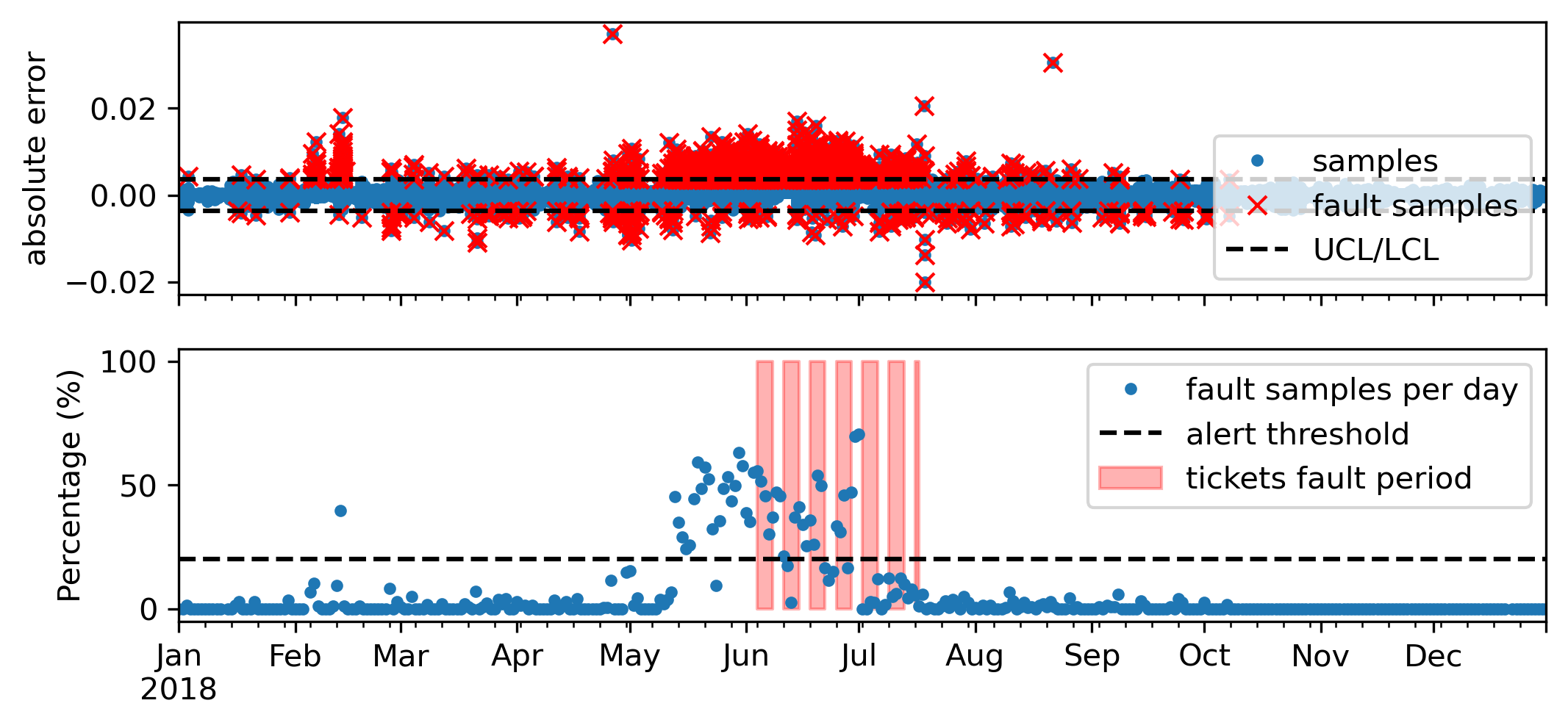}
    \caption{Example of statistical analysis using Shewhart chart for monitoring the absolute deviation of \textit{5min single} measurements. Simulated output using ARX modeling method.}
    \label{fig:example_shewhart_absolute_5min_ARX}
\end{figure}

\begin{figure}[hp]
    \centering
    \includegraphics[width=13cm]{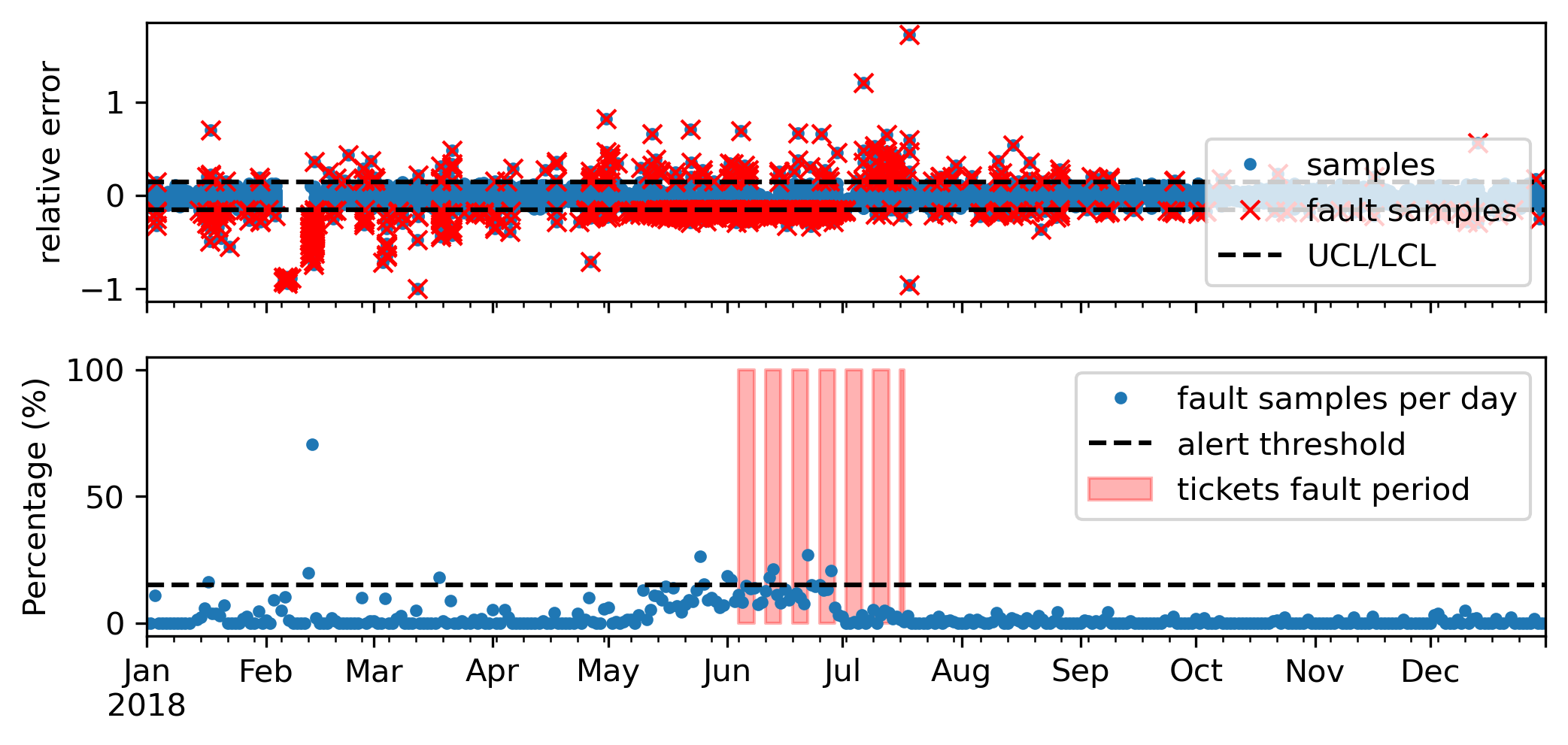}
    \caption{Example of statistical analysis using Shewhart chart for monitoring the relative deviation of \textit{5min single} measurements. Simulated output using ARX modeling method.}
    \label{fig:example_shewhart_relative_5min_ARX}
\end{figure}

The use of absolute deviation in comparison to the relative deviation resulted in better performance, specially in terms of weighted sensitivity. Figure~\ref{fig:example_shewhart_absolute_5min_ARX} and Figure~\ref{fig:example_shewhart_relative_5min_ARX} show examples of \textit{absolute} and \textit{relative} deviations, respectively, analyzed individually (\textit{5 min single}) through \textit{Shewhart} charts. In both figures, the upper plot depicts the chart with the deviation values outside the control limits marked in red 'x', while the lower plot shows the percentage of time "out of control" in relation to the daylight time. The red shades observed in those plots represents the faulty period according the maintenance tickets. Interestingly, the high level of deviation appears days before the fault is detected by the conventional monitoring routine and recorded in the maintenance logs.

\begin{figure}[hp]
    \centering
    \includegraphics[width=13cm]{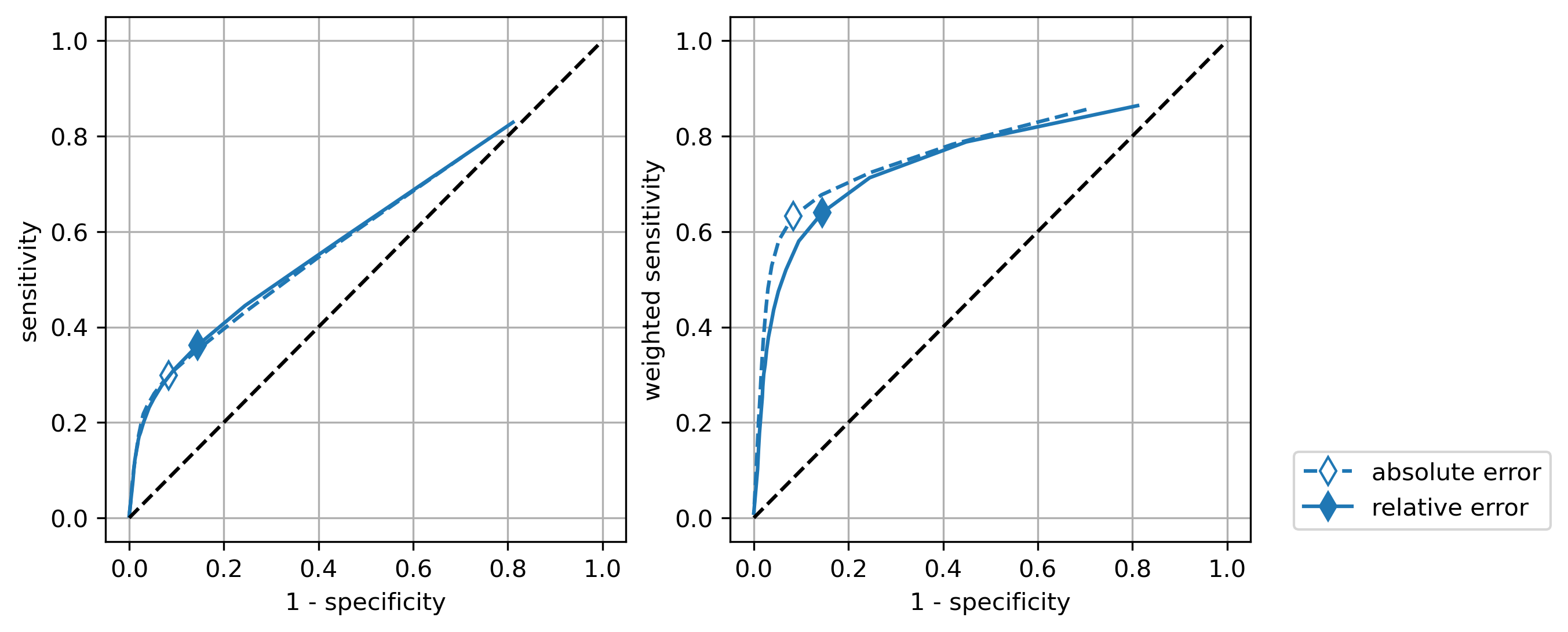}
    \caption{The ROC curve for the AFD algorithms based on statistical analysis using Shewhart chart monitoring absolute and relative deviation of \textit{5min single} measurements. Simulated output using ARX modeling method.}    
    \label{fig:example_ROC_shewhart_5min_ARX}
\end{figure}

Large percentages are most probably related to fault events, thus a fault alert is triggered when the it crosses a threshold. The optimum threshold identified according to the Youden's method using the ROC curve is shown in Figure~\ref{fig:example_ROC_shewhart_5min_ARX}. 
In the lower plots of Figures \ref{fig:example_shewhart_absolute_5min_ARX} and \ref{fig:example_shewhart_relative_5min_ARX}, we can see that the changes in the relative deviation are more subtle than in the absolute deviation. This can make it difficult to distinguish between faulty and non-faulty days, resulting in more false alerts and lower specificity. 

\begin{figure}[hp]
    \centering
    \includegraphics[width=13cm]{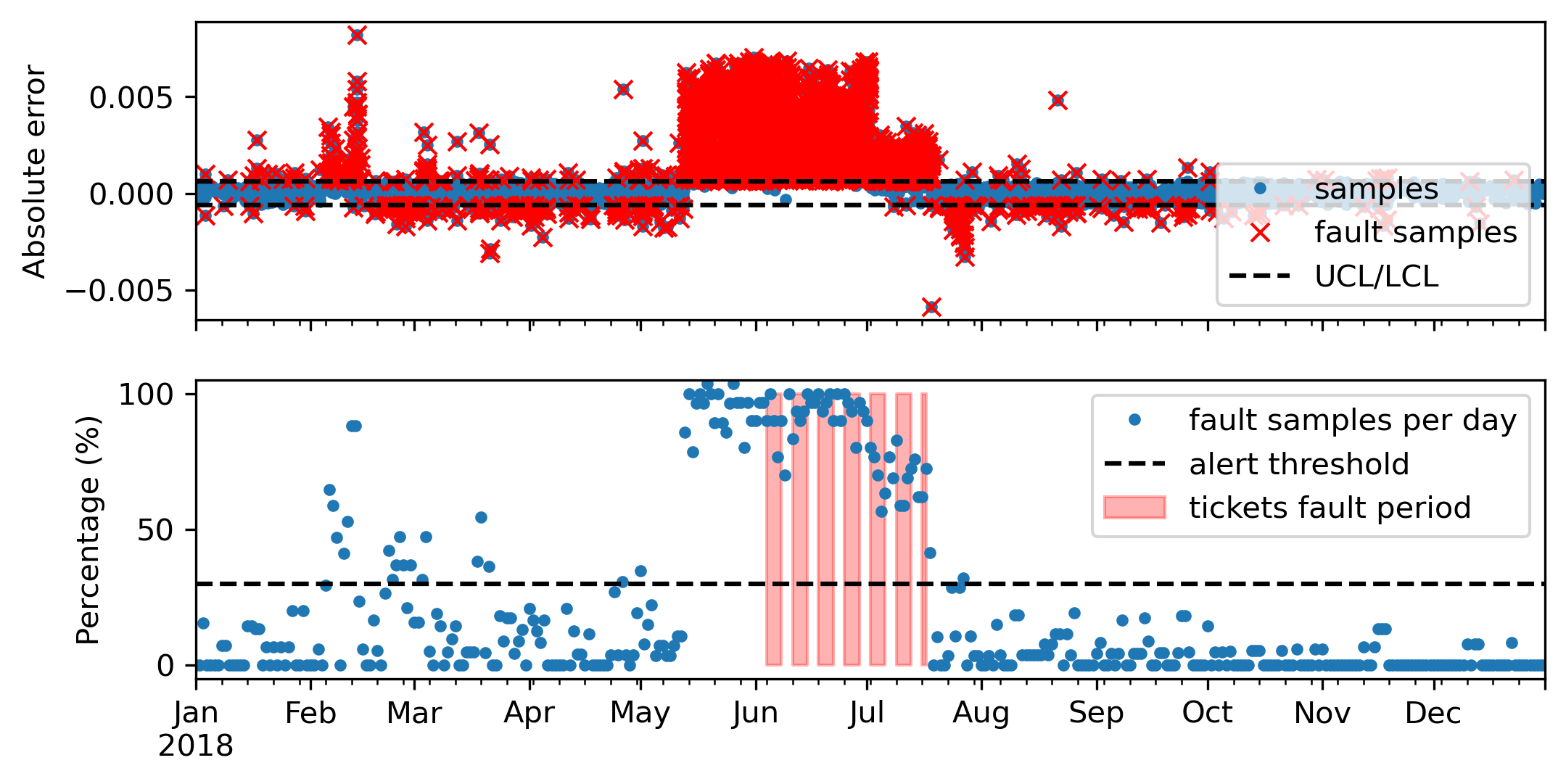}
    \caption{Example of statistical analysis using Shewhart chart for monitoring the average absolute deviation of measurements grouped every 30 minutes. Simulated output using ARX modeling method.}
    \label{fig:example_shewhart_absolute_30min_ARX}
\end{figure}

Although there is a difference in performance between the two types of deviations, the influence of the grouping method is greater.
In Figure~\ref{fig:ROC_shewhart_ARX_all} we observe that gathering the samples into small groups increased the sensitivity, but reduced the specificity slightly.
An example of \textit{Shewhart} charts for analyzing the measurements every \textit{30 min} is depicted in Figure~\ref{fig:example_shewhart_absolute_30min_ARX}. It shows a large dispersion in the percentage of fault samples, which can help distinct fault events. But it also amplifies the noise during non-fault periods.

\begin{figure}[hp]
    \centering
    \includegraphics[width=13cm]{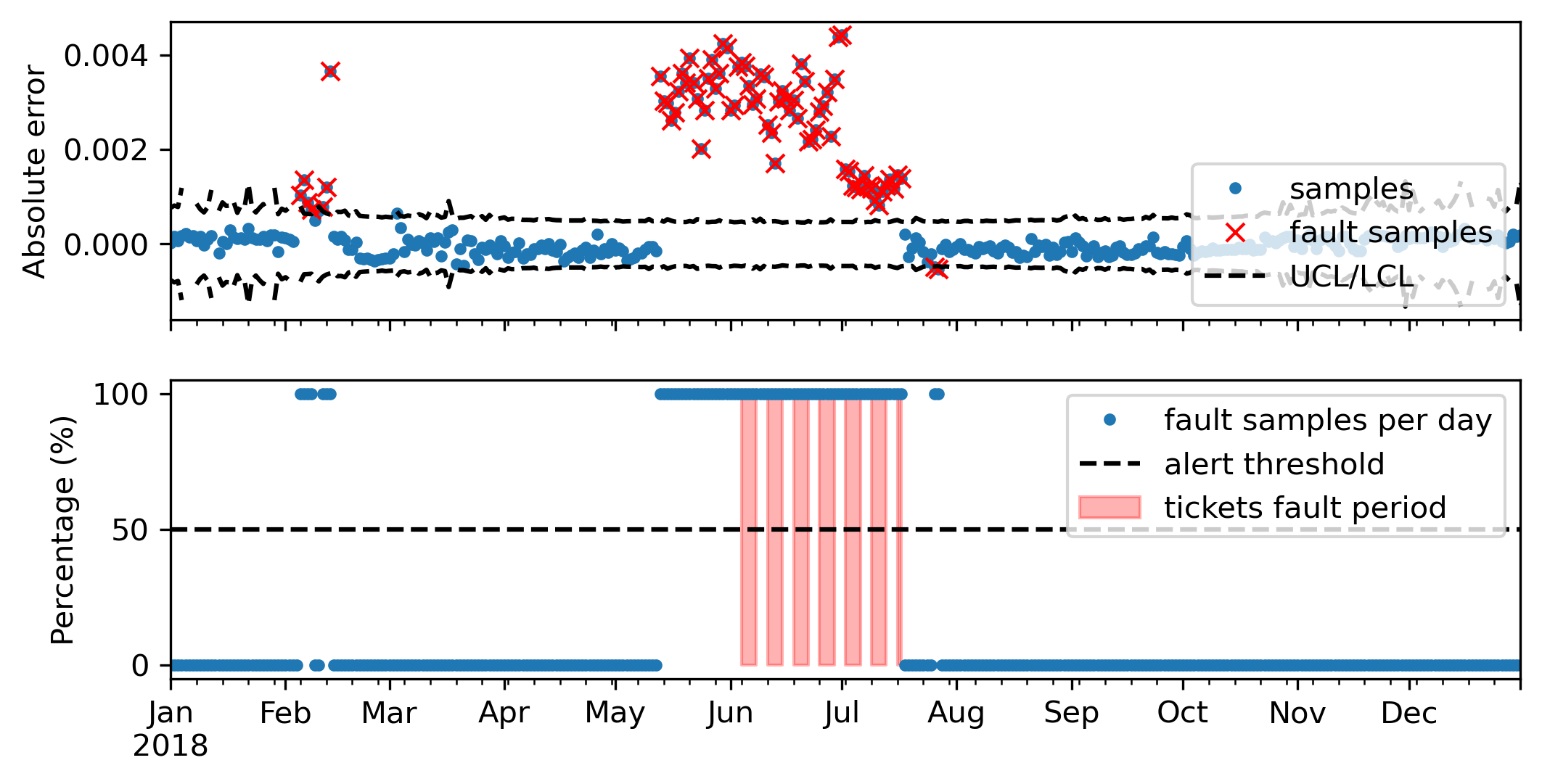}
    \caption{Example of statistical analysis using Shewhart chart for monitoring the average absolute deviation of measurements grouped every day. Simulated output using ARX modeling method.}
    \label{fig:example_shewhart_absolute_dailygroup_ARX}
\end{figure}

Using \textit{daily groups} resulted in even lower specificity. In this case, there is only one sample to evaluate per day, thus the chart analysis provides daily Boolean results (within control limits or not) leaving no chance to refine the alerts by tuning the threshold. Also, daily groups have sample size $n$ that varies throughout the year according to the length of the day. This results in varying control limits, as can be observed in the example of \textit{Shewhart} charts shown in Figure~\ref{fig:example_shewhart_absolute_dailygroup_ARX}.

\begin{figure}[hp]
    \centering
    \includegraphics[width=13cm]{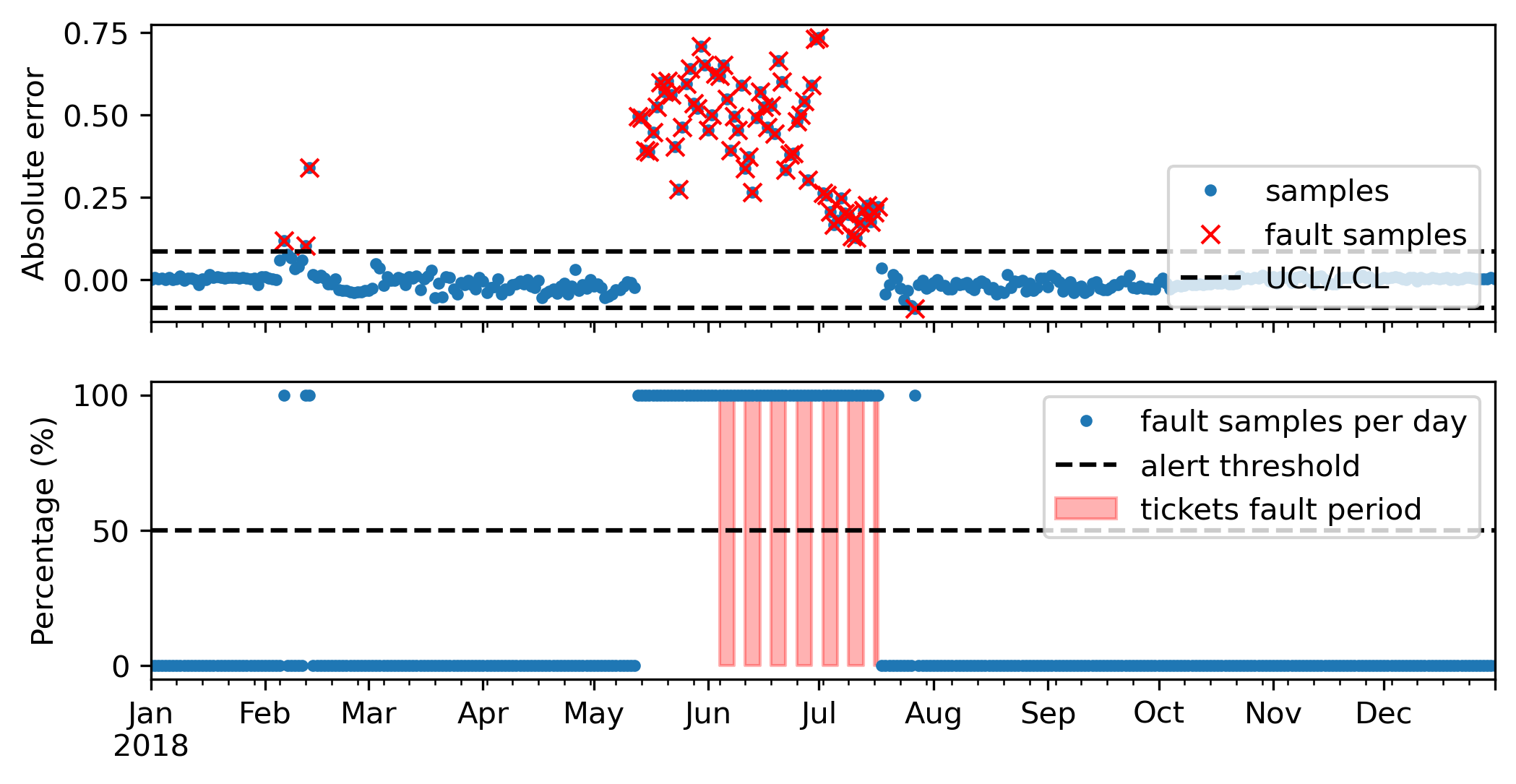}
    \caption{Example of statistical analysis using Shewhart chart for monitoring the daily absolute deviation. Predicted output estimated using ARX modeling method.}
    \label{fig:example_shewhart_absolute_dailysingle_ARX}
\end{figure}

Aggregating the measurements into a single deviation value per day improved the sensitivity, but the specificity is still not so good. There is also no chance to refine the alerts by adjusting the threshold, but the graphs for single values show a wider control threshold (due to smaller $n$), which reduces false alerts. See an example in Figure~\ref{fig:example_shewhart_absolute_dailysingle_ARX}.

\begin{figure}[hp]
    \centering
    \includegraphics[width=13cm]{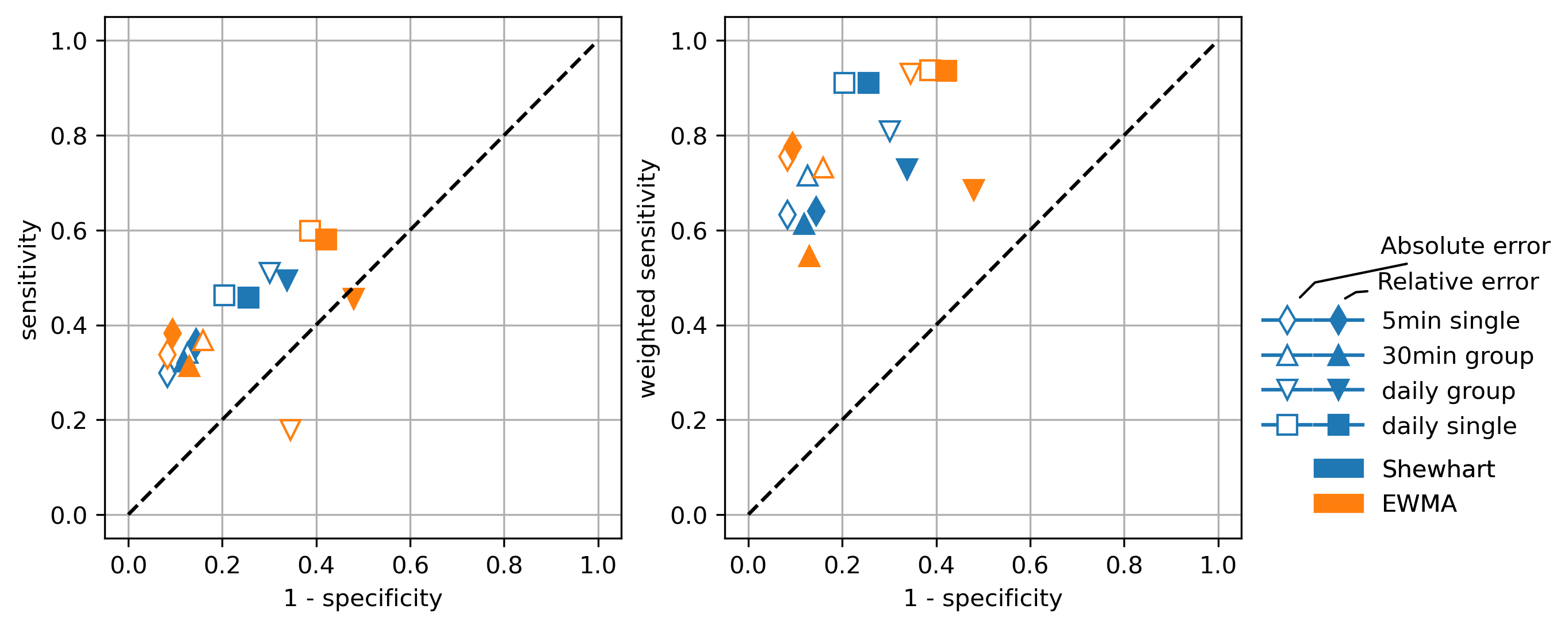}
    \caption{Sensitivity and specificity for all Shewhart and EWMA charts with ARX modeling.}
    \label{fig:ROC_EWMA_ARX_all}
\end{figure}

\begin{figure}[hp]
    \centering
    \includegraphics[width=13cm]{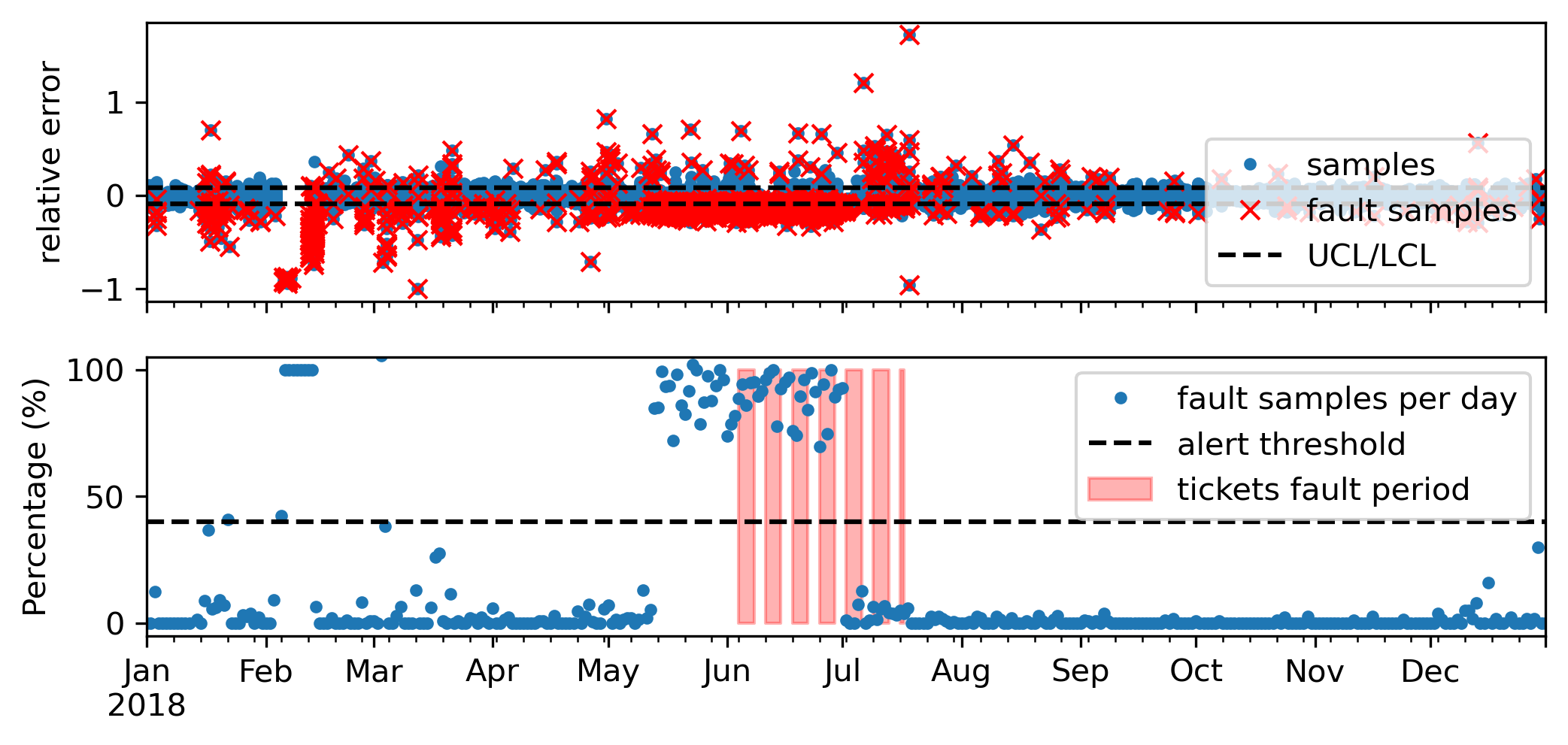}
    \caption{Example of statistical analysis using EWMA chart for monitoring the relative deviation of individual measurements. Simulated output using ARX modeling method.}
    \label{fig:example_EWMA_relative_5min_ARX}
\end{figure}

\begin{figure}[hp]
    \centering
    \includegraphics[width=13cm]{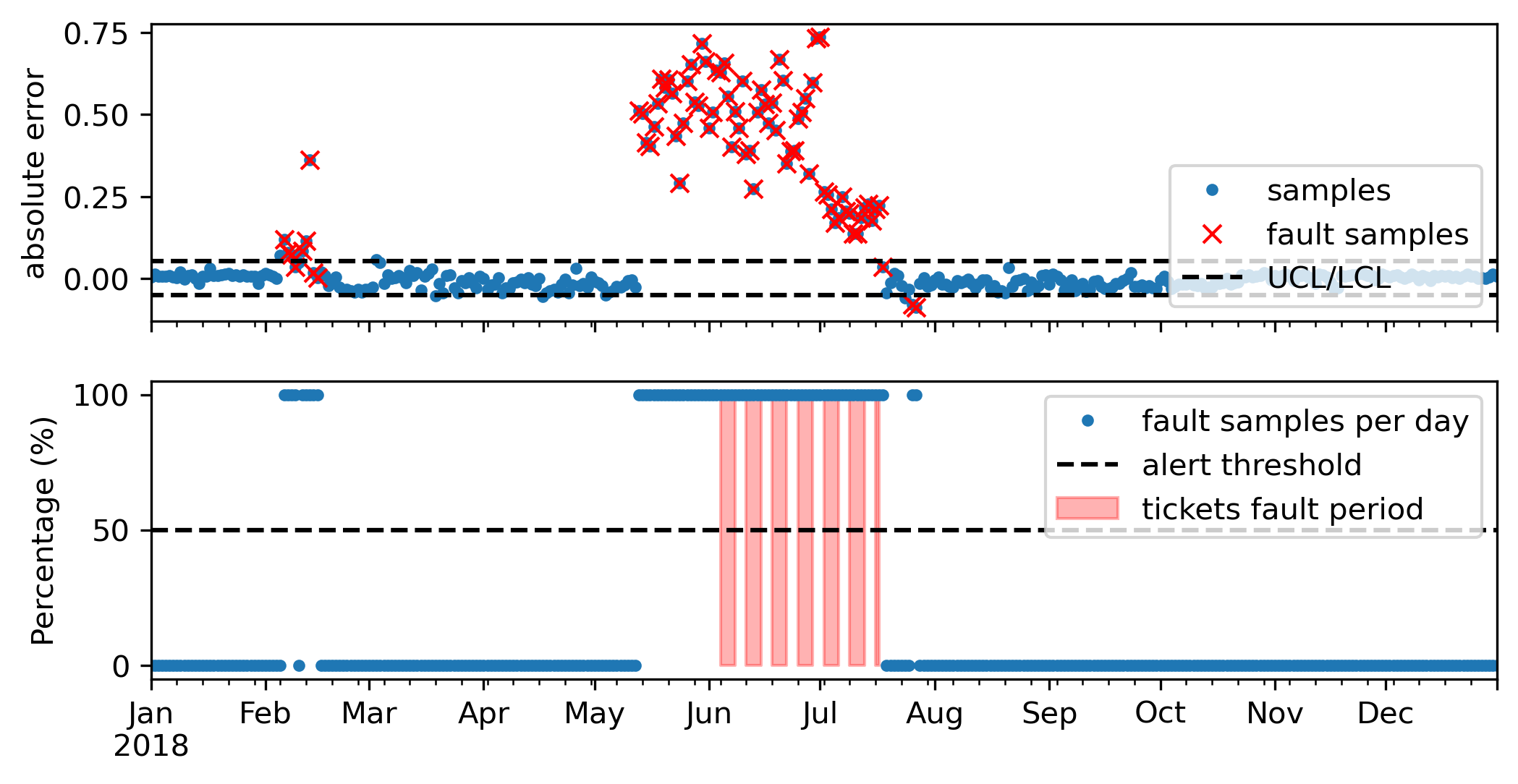}
    \caption{Example of statistical analysis using EWMA chart for monitoring the daily absolute deviation. Simulated output using ARX modeling method.}
    \label{fig:example_EWMA_absolute_dailysingle_ARX}
\end{figure}

Similar AFD algorithms, but using the EWMA chart instead of the Shewhart chart, were tested and the resulting sensitivities and specificities are summarized in Figure~\ref{fig:ROC_EWMA_ARX_all}.
Overall, the change of chart type to EWMA had a negative impact, except for the \textit{5 min single} analysis, where the weighted sensitivity increased by 13\% without affecting the specificity. One example of this analysis depicted in Figure~\ref{fig:example_EWMA_relative_5min_ARX} shows that the EWMA chart increased the difference between normal and faulty days, which explains the better fault detection performance. In the other cases, the EWMA caused a reduction in specificity possibly due to the reduction of the tolerance window according to the smoothing factor, which resulted in more false alerts. See an example in Figure~\ref{fig:example_EWMA_absolute_dailysingle_ARX}.

\subsection{The impact of modeling accuracy} \label{sec:impact_modeling}

\begin{figure}[h!]
    \centering
    \includegraphics[width=13cm]{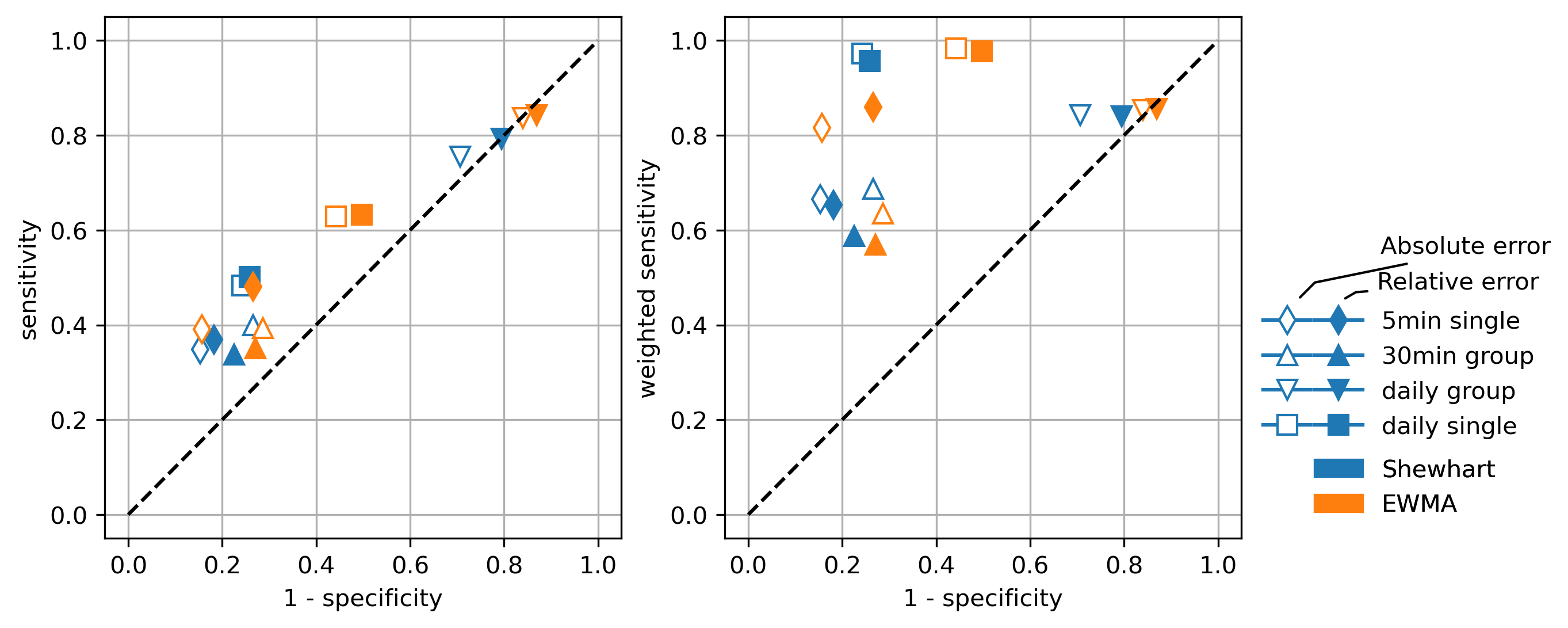}
    \caption{Sensitivity and specificity for all Shewhart and EWMA charts with Polynomial Regression modeling method.}
    \label{fig:ROC_EWMA_PolyReg_all}
\end{figure}

\begin{figure}[h!]
    \centering
    \includegraphics[width=13cm]{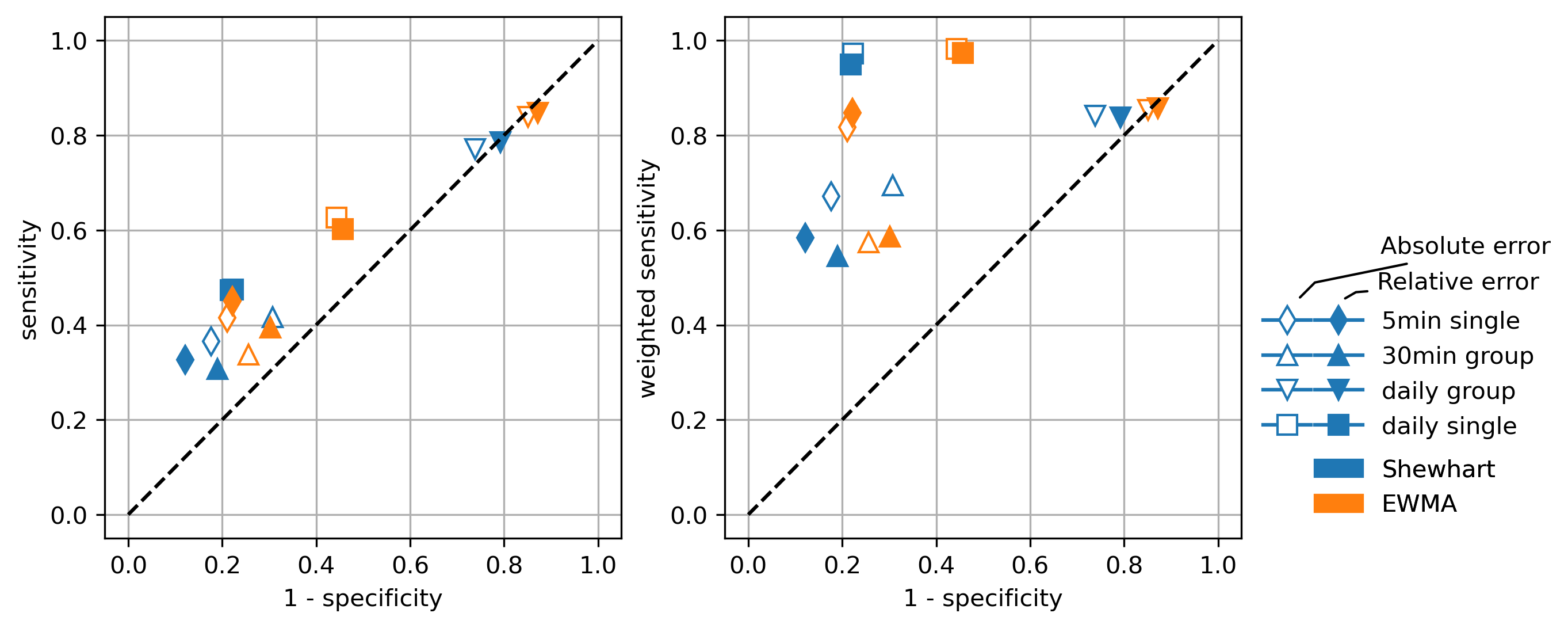}
    \caption{Sensitivity and specificity for all Shewhart and EWMA charts with empirical modeling method.}
    \label{fig:ROC_EWMA_Empirical_all}
\end{figure}

After testing the AFD algorithms using the most accurate modeling method, we repeated all tests using less accurate modeling methods to investigate the impact of the simulation accuracy in the fault detection process. 
Figure~\ref{fig:ROC_EWMA_PolyReg_all} and Figure~\ref{fig:ROC_EWMA_Empirical_all} show the resulting performances (sensitivities and specificities) using the simulation with polynomial regression and empirical model, respectively. 
In general, the use of a less accurate model resulted in lower specificity.
The decrease is lower for the analysis with single measurements (\textit{5 min} or \textit{daily}) and stronger for the rational groups (\textit{30 min} or \textit{daily}), especially the daily grouping. 
Regarding the sensitivity, the effect of less accurate modeling was positive in some cases (\textit{5 min single}, and \textit{daily single}), and neutral to slightly negative in others. It can be said that grouping samples can make statistical analysis more sensitive to modeling accuracy. Analyzing measurements individually results in more robust AFD algorithms.

\begin{figure}[h!]
    \centering
    \includegraphics[width=13cm]{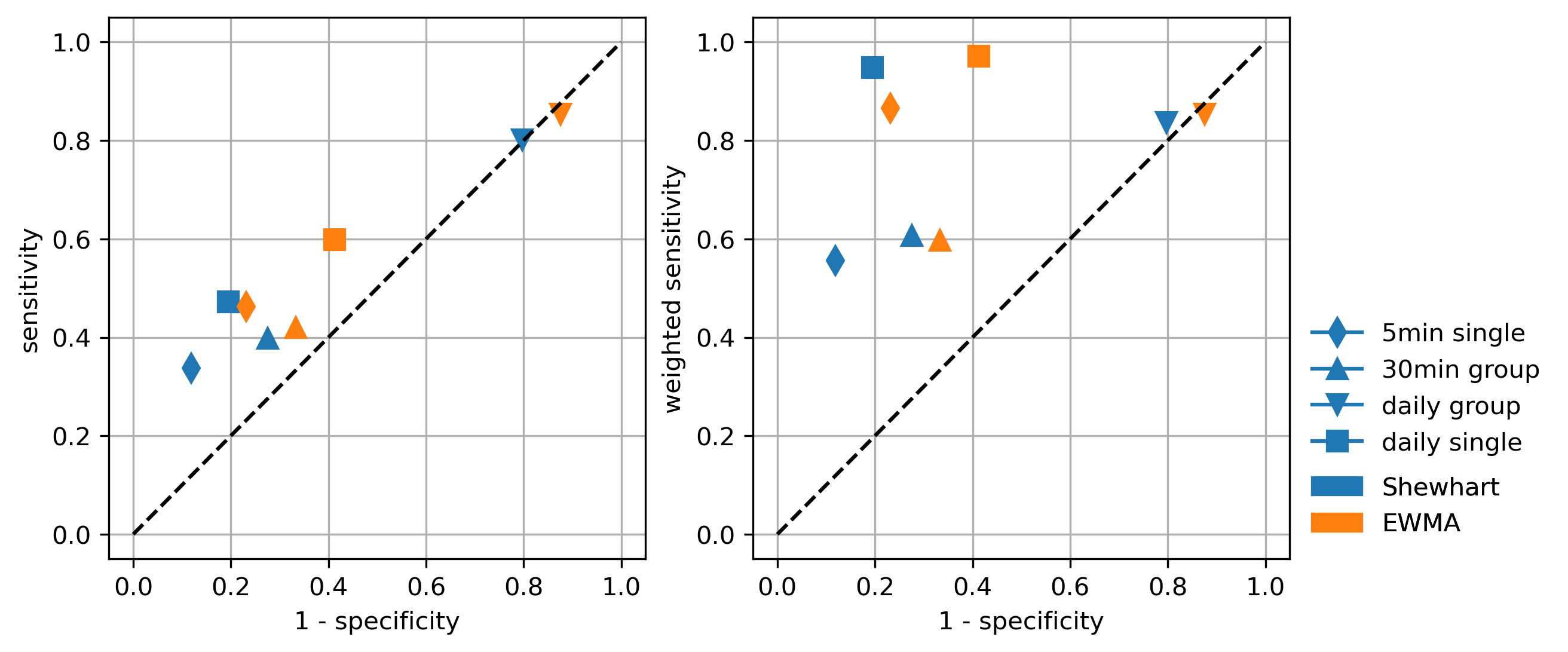}
    \caption{Sensitivity and specificity for all Shewhart and EWMA charts with PR.}
    \label{fig:ROC_shewhart_EWMA_PR}
\end{figure}

\begin{figure}[h!]
    \centering
    \includegraphics[width=13cm]{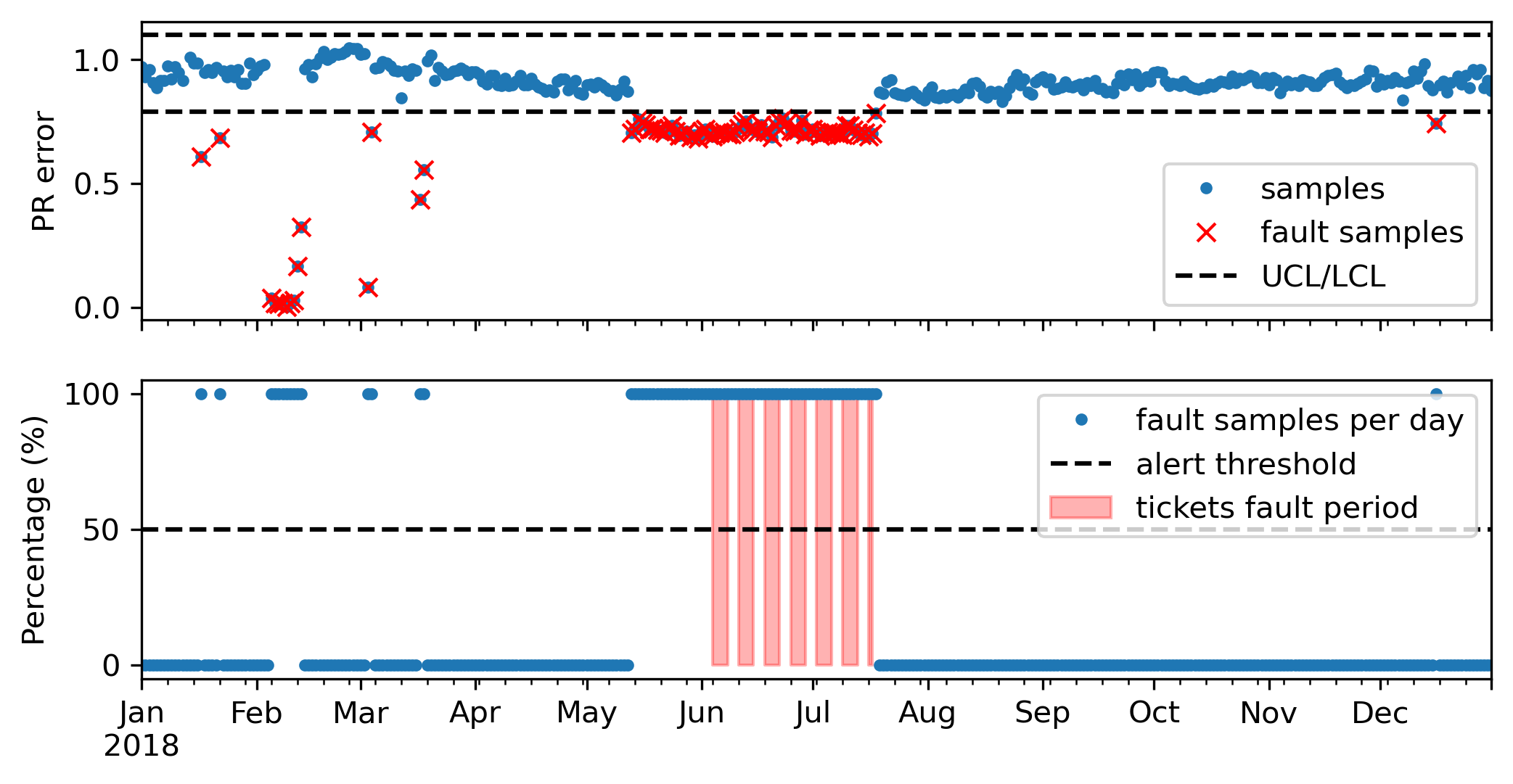}
    \caption{Example of statistical analysis using Shewhart chart for monitoring the daily PR.}
    \label{fig:example_shewhart_PR}
\end{figure}

Suppose no modeling tool is available for the monitoring of the PV systems.
In this case, it is common to use the performance ratio (PR) to evaluate the performance of the system. Here we tested some AFD algorithms adopting control charts to monitor the PR calculated every \textit{5 min} or \textit{daily}. Figure~\ref{fig:ROC_shewhart_EWMA_PR} summarizes the sensitivities and specificities obtained. The results showed a performance very similar to the AFD algorithms with low accuracy modeling, being the best performance achieved by the algorithm analyzing the daily PR with Shewhart chart. See one example of the daily PR monitoring with Shewhart chart in Figure~\ref{fig:example_shewhart_PR}.

\subsection{Using a machine learning tool for the statistical analysis}

\begin{figure}[h!]
    \centering
    \includegraphics[width=13cm]{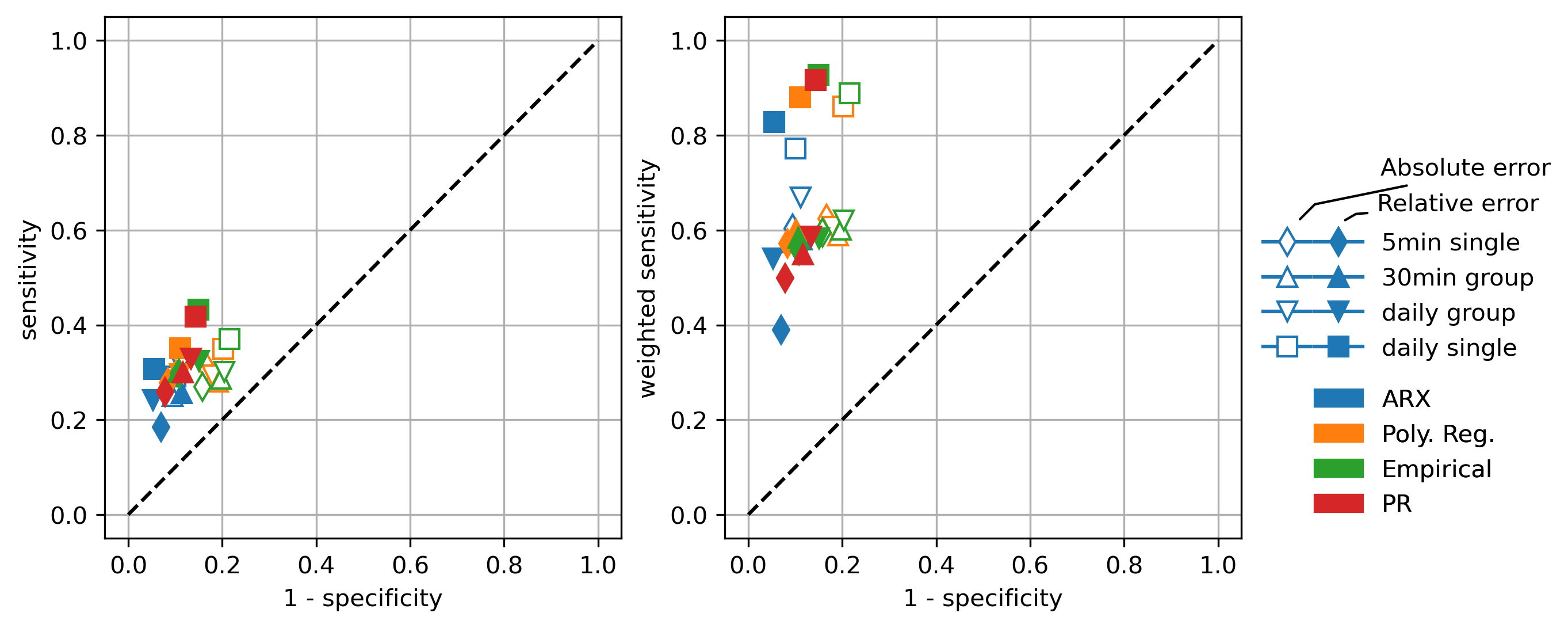}
    \caption{Sensitivity and specificity for all AFD algorithms using K-means clustering for the statistical analysis.}
    \label{fig:ROC_kmeans_all}
\end{figure}

\begin{figure}[h!]
    \centering
    \includegraphics[width=13cm]{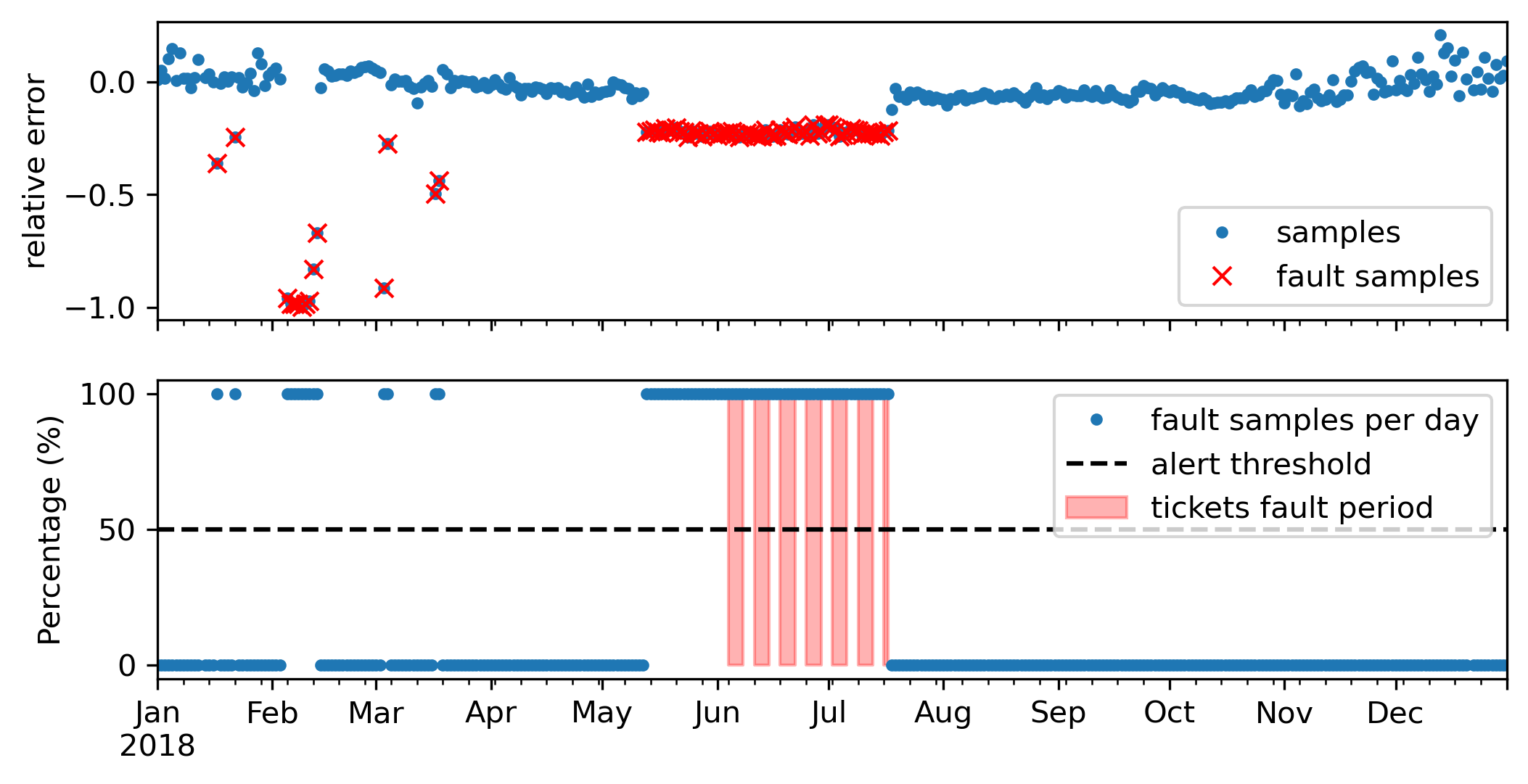}
    \caption{Example of statistical analysis using K-means algorithm for monitoring the daily relative deviation based on the simulation with polynomial regression.}
    \label{fig:example_kmeans_dailysingle_relative_PolyReg}
\end{figure}

The performances of the AFD algorithms using the proposed statistical analysis based on k-means clustering are summarized in Figure~\ref{fig:ROC_kmeans_all}. The specificity values obtained indicate that the proposed algorithm is very robust in terms of false alerts. None of the cases obtained a specificity below 78\%, and most cases are around 90\%. The algorithm showed similar performance for all methods using 5 min measurements (\textit{5min single}, \textit{30min group} or \textit{daily group}), with sensitivity around 55\% and specificity around 90\%. The best performance was seen with \textit{daily single} \textit{relative} deviations. See one example in Figure~\ref{fig:example_kmeans_dailysingle_relative_PolyReg}.

\subsection{Overview of the results and summary of best methods}

The results show that none of the tested AFD algorithms was capable of detecting all faults recorded in the maintenance tickets. The best performing algorithms were able to identify up to 38.3\% of the faults while maintaining specificity above 90\%. Nonetheless, some AFD algorithms detected the faults associated with more than 82.8\% of the total energy losses observed in those systems. 
This confirms the hypothesis that the maintenance records are not necessarily associated with problems that significantly affect system performance. Any AFD algorithm based on the difference between measured and simulated output power will have trouble detecting those type of faults. Considering this, the discussion of the tested AFD algorithms is mainly based on weighted sensitivity.

The increase of the sensitivity level is generally associated with a decrease of the specificity, i.e., an increasing the occurrence of false alerts. A large number of false alerts is not tolerable in the monitoring process of large portfolios, so achieving high specificity is critical for an AFD algorithm. 
Maintenance records are not a perfect reference, they reflect the actions of inspection and maintenance personnel. Thus, part of the false alerts may actually be unrecorded faults that were detected by the AFD algorithm. Nonetheless, maintenance records remain the best record of the plant's history and the best available reference for measuring the quality of AFD algorithms.

The performance indexes for the AFD algorithms which achieved \textit{Specificity} higher than 80\% and \textit{Weighted Sensitivity} higher than 75\% are summarized in Table~\ref{tab:best_algorithms}. Those are considered the best of all AFD algorithms tested. 

Note that none of them analyze the samples in groups, the best analyses are \textit{5 min single} or \textit{daily single}. \textit{5 min single} samples are best analyzed with EWMA plots, while \textit{daily single} samples perform better with k-means clustering. In general, the higher the accuracy of the model, the higher the specificity. The use of less accurate modeling, or simply PR, can increase sensitivity at the cost of decreasing specificity.

With the lowest specificity among the best algorithms, the analysis of daily PR using a Shewhart chart provides the highest sensitivity with an exceptionally simple solution with no need for more complex algorithms for modeling or clustering.

\begin{table}[h]
\centering
\caption{Summary of best AFD algorithms tested sorted by their \textit{Specificity}.}
\label{tab:best_algorithms}
\small
\begin{tabular}{ccccccc}
\hline
\multicolumn{1}{c}{\begin{tabular}[c]{@{}c@{}}Statistical\\ Analysis\end{tabular}} &
  \multicolumn{1}{c}{Modeling} &
  \multicolumn{1}{c}{Grouping} &
  \multicolumn{1}{c}{Deviation} &
  \multicolumn{1}{c}{Sensitivity} &
  \multicolumn{1}{c}{\begin{tabular}[c]{@{}c@{}}Weighted\\ Sensitivity\end{tabular}} &
  \multicolumn{1}{c}{Specificity} \\ \hline
k-means  & ARX         & daily single & relative & 0.307 & 0.828 & 0.945 \\
EWMA     & ARX         & 5 min single & absolute & 0.338 & 0.755 & 0.917 \\
EWMA     & ARX         & 5 min single & relative & 0.383 & 0.776 & 0.905 \\
k-means  & ARX         & daily single & absolute & 0.291 & 0.772 & 0.899 \\
k-means  & PolyReg     & daily single & relative & 0.351 & 0.880 & 0.890 \\
k-means  & -           & daily single & PR       & 0.418 & 0.917 & 0.857 \\
k-means  & Empirical   & daily single & relative & 0.432 & 0.928 & 0.851 \\
EWMA     & PolyReg     & 5 min single & absolute & 0.391 & 0.816 & 0.843 \\
Shewhart & -           & daily single & PR       & 0.472 & 0.948 & 0.805 \\ \hline
\end{tabular}%
\end{table}

\section{Conclusion}

A series of AFD algorithms was tested using field data from a portfolio of 80 commercially operated PV power plants located in Germany. Those AFD algorithms are based on the comparison between measured and simulated PV system behavior. The influence of various aspects regarding the design of the AFD algorithm have been investigated and practical recommendations have been made based on comparisons of the algorithms' performances. The best performing algorithms were able to identify up to 38.3\% of the faults and 82.8\% of the energy losses with specificity above 90\%. 

Implementing any of the AFD algorithms tested here is inexpensive since it does not require the installation of additional sensors beyond the common monitoring equipment already available in commercially operated PV systems. The information required for adapting the algorithms for different PV systems can be obtained from the system planning (e.g. equipment’s rated power) or extracted directly from the measurements (e.g. the average and standard deviation). At least one year of measurements might be needed for tuning the algorithm’ parameters.

Fault detection alerts with daily resolution are not sufficient for monitoring within O\&M contracts with strict specifications, e.g. reaction time within a few hours. Considering this scenario, future work will be dedicated to study hourly alerts. To this end, high precision simulation based on physical modeling chain (like Fraunhofer Zenit\texttrademark) will be tested to implement a suitable AFD algorithm.

\section{Acknowledgments}

The authors would like to thank all co-workers at Fraunhofer ISE, Pohlen Solar and at IDMEC for supporting this work. This work was partly supported by FCT, through IDMEC, under LAETA, project UID/EMS/50022/2020. This work was partly financed by the Federal Ministry for Economic Affairs and Energy of Germany (BMWi) within the framework of the research program "Innovations for the energy transition" funded under the contract number 03EE1058.

\bibliographystyle{elsarticle-num} 
\bibliography{references.bib}



\end{document}